\documentclass[a4paper,11pt]{article}
\pdfoutput=1

\usepackage{jinstpub}
\usepackage{lineno}
\graphicspath{{figure/}}

\title{Design and commissioning of a 600~L Time Projection Chamber with Microbulk Micromegas}

\author[a]{Heng Lin,}
\author[b]{Denis Calvet,}
\author[c]{Lei Chen,}
\author[a]{Xun Chen,}
\author[d]{Theopisti Dafni,}
\author[a]{Changbo Fu,}
\author[a,b]{Javier Galan,}
\author[a,1]{Ke Han,\note{Corresponding authors.}}
\author[c]{Shouyang Hu,}
\author[a]{Yikai Huo,}
\author[d]{Igor G. Irastorza,}
\author[a,e]{Xiangdong Ji,}
\author[c]{Xiaomei Li,}
\author[c]{Xinglong Li,}
\author[a]{Jianglai Liu,}
\author[a,d]{Hector Mirallas,}
\author[b]{Damien Neyret,}
\author[a]{Kaixiang Ni,}
\author[f]{Hao Qiao,}
\author[a]{Xiangxiang Ren,}
\author[a,g,1]{Shaobo Wang,}
\author[f]{Siguang Wang,}
\author[a]{Yong Yang,}
\author[f]{Ying Yuan,}
\author[a]{Tao Zhang}
\author[a]{and Li Zhao}

\affiliation[a]{INPAC and School of Physics and Astronomy, Shanghai Jiao Tong University, Shanghai Laboratory for Particle Physics and Cosmology, 800 Dongchuan Rd., Shanghai 200240, China}
\affiliation[b]{IRFU, CEA, Universit\'e Paris-Saclay, F-91191 Gif-sur-Yvette, France}
\affiliation[c]{Science and Technology on Nuclear Data Laboratory, China Institute of Atomic Energy, 1 Sanqiang Rd., Beijing 102413, China}
\affiliation[d]{Grupo de F\'isica Nuclear y Astropart\'iculas, Departamento de F\'isica Te\'orica, Universidad de Zaragoza, C/ Pedro Cerbuna 12, 50009, Zaragoza, Spain}
\affiliation[e]{Department of Physics, University of Maryland, 82 Regents Dr., College Park, MD 20742, U.S.A}
\affiliation[f]{School of Physics, Peking University, 5 Yiheyuan Rd., Beijing 100871, China}
\affiliation[g]{SJTU-Paris Tech Elite Institute of Technology, Shanghai Jiao Tong University, 800 Dongchuan Rd., Shanghai 200240, China}

\emailAdd{ke.han@sjtu.edu.cn; shaobo.wang@sjtu.edu.cn}

\abstract{
We report the design, construction, and initial commissioning results of a large high pressure gaseous Time Projection Chamber (TPC) with Micromegas modules for charge readout.
The detector vessel has an inner volume of about 600~L and an active volume of 270~L.
At 10~bar operating pressure, the active volume contains about 20~kg of xenon gas and can image charged particle tracks.
Drift electrons are collected by the charge readout plane, which accommodates a tessellation of seven Micromegas modules.
Each of the Micromegas covers a square of 20~cm by 20~cm.
A new type of Microbulk Micromegas is chosen for this application due to its good gain uniformity and low radioactive contamination.
Initial commissioning results with 1 Micromegas module running with 1~bar argon and isobutane gas mixture and 5~bar xenon and trimethylamine (TMA) gas mixture are reported.
We also recorded extended background tracks from cosmic ray events and highlighted the unique tracking feature of this gaseous TPC.
}


\arxivnumber{1804.02863}

\begin{document}
\maketitle
\flushbottom

\section{Introduction}
\label{sec:intro}

Time Projection Chambers (TPC) can image three-dimensional ionization trajectories of ionizing particles travelling within, and are widely used in particle physics experiments since the invention in 1970s~\cite{Nygren:1978rx}.
More recently, TPC's application is expanded to rare event searches, such as neutrinoless double beta decay experiments (e.g., NEXT~\cite{Martin-Albo:2015rhw} and EXO-200~\cite{Albert:2014awa}), dark matter WIMP direct detection experiments (e.g., PandaX-II~\cite{Tan:2016zwf}, LUX~\cite{Akerib:2016vxi}, XENON1T~\cite{Fieguth:2017aca} and TREX-DM~\cite{Iguaz:2015myh}) and directional detection of galactic WIMP signals (e.g., DRIFT~\cite{Daw:2010ud} and MIMAC~\cite{Santos:2013oua}).
A high pressure gaseous TPC combines attractive features of 3D imaging, excellent intrinsic energy resolution~\cite{BOLOTNIKOV1997360}, and potentially large target mass.
PandaX-III ~\cite{Chen:2016qcd} is a newly proposed experiment to utilize High Pressure Xenon (HPXe) gaseous TPC to search for neutrinoless double beta decay (NLDBD) of $^{136}$Xe.
For NLDBD searches, a gaseous TPC provides good energy resolution at the decay Q-value and more importantly event tracks to greatly improve the background reduction capability in the region of interest.
In our conceptual design, the PandaX-III TPC has an inner volume of 4~m$^3$ and can hold 200~kg of xenon at 10~bar operating pressure.
In contrast with other proposed experiments, only ionization signals from the interaction between the target particle and the gas medium would be recorded in PandaX-III.
To study key features of HPXe TPC, especially the tracking and energy response, we have designed, constructed, and commissioned a 600~L gaseous TPC, which is the main topic of this paper.
The detector has an active volume with a 78~cm drift distance and a 66~cm diameter (about 270~L), large enough to contain MeV-scale electron tracks with high pressure xenon gas.
It has about one seventh of the volume of the PandaX-III 200~kg detector, and therefore is technically representative.

The charge readout plane consists of a tessellation of seven 20$\times$20~cm$^2$ Micromegas readout modules.
Micromegas, short for \emph{Micro-Mesh Gaseous Structure}, is a successful Micro Pattern Gas Detector (MPGD) technique widely used in particle and nuclear physics detectors~\cite{Giomataris:1995fq}.
The Micromegas we use is fabricated with the so-called \emph{Microbulk} technology~\cite{Andriamonje:2010zz}.
The main advantages of this technology reside in its good energy resolution, good spatial resolution, and the radiopurity of the materials that these readouts are made of.
For those reasons, Microbulk Micromegas has been selected as the readout technology for the first phase of PandaX-III experiment.
A strip readout scheme, where pixels with the same horizontal or vertical coordinates are connected in series and read out together, is chosen to reduce the number of readout channels.
Our work is one of the first and largest detectors aiming to characterize the performance of Microbulk Micromegas with strip readout~\cite{Irastorza:2015dcb,Gonzalez-Diaz:2015oba}.

We performed commissioning runs with 1 active Micromegas at the center of the charge readout plane and with different gas mixtures.
Argon gas mixed with 5\% isobutane was used first at 1~bar pressure to demonstrate the basic operation of the Micromegas detector and the data acquisition system.
Xenon gas with 1\% of trimethylamine (TMA) mixture is the baseline choice for PandaX-III TPC and we tested the detector performance at 5~bar.
Attributed to Penning effect, the TMA in xenon improves the detector gain and resolution~\cite{Cebrian:2012sp}.
A $^{241}$Am calibration source was used and we ran extensive Monte Carlo simulations to help understand the measured low energy spectrum.

The structure of this paper is as follows.
The experimental setups, such as the vessel, field cage, feedthroughs, gas system and DAQ system are presented in Section~\ref{sec:experimental setup}.
The charge readout plane of Micromegas modules is described in details in Section~\ref{sec:MM}.
Then simulation result and the first performance in Ar-(5\%)isobutane mixture and Xe-(1\%)TMA mixture, including examples of tracks and energy resolution are summarized in Section~\ref{sec:data-taking}.
Conclusions and outlook are presented in Section~\ref{sec:outlook}.

% !TEX root = P3_Prototype_1MM.tex

\section{Time Projection Chamber and Readout Electronics}
\label{sec:experimental setup}

\begin{figure}[tb]
  \begin{center}
    \includegraphics[width=\linewidth]{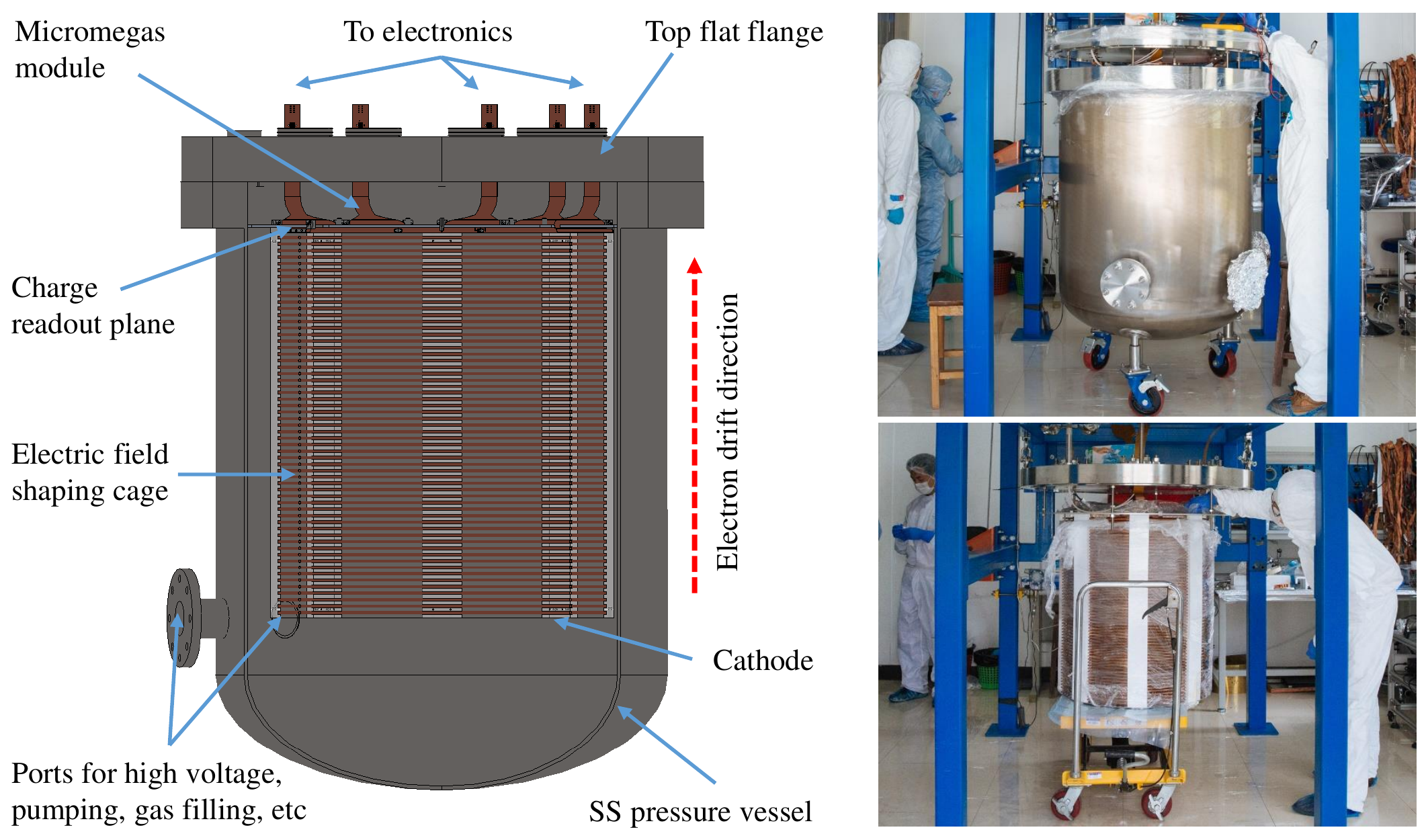}
    \caption{
    (Left): Cutaway drawing of the TPC and the pressure vessel with main components highlighted. Signals from 7 Micromegas (with 5 shown in the picture) are connected to front end electronics (not shown) for readout. There are four side ports on the lower bottom of the vessel while only two are visible in the drawing. (Top right): The SS pressure vessel during assembly. (Bottom right): Mounting the TPC on to the top flat flange using threaded rods. During assembly, the top flat flange and TPC hang from a blue frame as seen in the pictures.
    }
    \label{fig:prototypeTPC}
  \end{center}
\end{figure}

\begin{figure}[tb]
  \begin{center}
    \includegraphics[width=\linewidth]{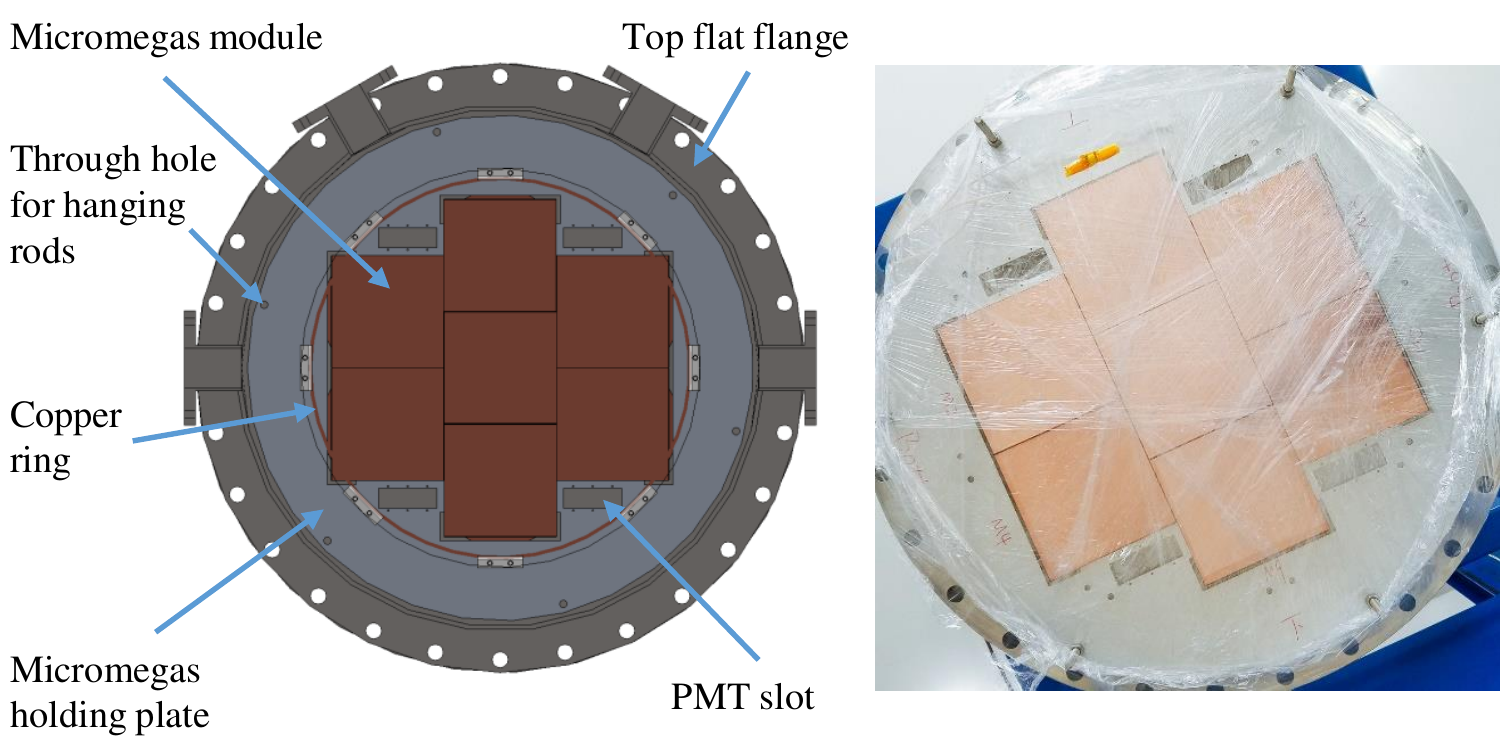}
    \caption{
   (Left): An illustration of the TPC from the bottom up to highlight the charge readout plane with Micromegas modules. Main components of the TPC and pressure vessel are annotated. (Right): A picture of the charge readout plane during assembly. To prevent dust and other contamination from attaching, the whole assembly is wrapped with polyethylene film, as visible in the picture. A small tube wrapped in Kapton contains a $^{241}$Am calibration source and is hanging from the charge readout plane.
   }
    \label{fig:chargeReadoutPlane}
  \end{center}
\end{figure}

The central part of our experimental setup is the time projection chamber itself.
It consists of a charge readout plane on the top, a cathode at the bottom and an electric field shaping cage in between, as shown in Fig.~\ref{fig:prototypeTPC}.
The active volume of the TPC is of cylindrical shape with a height of 78~cm, a diameter of 66~cm, and thus a total volume of about 270~L.
The TPC is hanging from the top flat flange of a 600~L stainless steel (SS) pressure vessel and the charge readout plane is kept parallel to the flange.
Detector signals from the charge readout plane are transmitted through the top flat flange via custom-made extension cable feedthroughs.
Signals are digitized and read out by a commercially available data acquisition suite based on AGET chips~\cite{ref:AGET_ASIC}.
On the lower side of the vessel, four small flanges provide ports for pumping, gas filling, and a high voltage feedthrough for the cathode.
A dedicated gas handling system is connected to the TPC vessel for gas filling, circulation, and purification.

While constructing the TPC, we paid special attention to the design and fabrication of individual parts as well as material selection to fulfill the requirement of vacuum and high gas pressure up to 10~bar.
On the other hand, we did not enforce the radioactive contamination screening since the main objective of this detector is to demonstrate the performance of large size HPXe TPC with Microbulk Micromegas.
The design and construction of each sub-system are described below.

\subsection{High pressure vessel}
\label{sec:prototype vessel}

The SS pressure vessel, with a design operating pressure of 15~bar, is shown in Fig.~\ref{fig:prototypeTPC}.
The top flange is flat for easier installation of the charge readout plane, and its 9~cm thickness provides enough mechanical strength.
The dome-shaped bottom allows thinner wall to withstand the same pressure.
The cylindrical side wall and bottom are both 0.8~cm thick.
The inner diameter of the cylindrical main body is 90~cm and the height is 100~cm.
The top flange is held onto the vessel with 28 M28 bolts.

Four DN-80 ports are at the lower part of the cylinder.
Besides a backup port, the other three are for the high voltage feedthrough of the cathode, gas inlet, and vacuum pumping.
The gas inlet port is equipped with a 1/4 inch SS pipe with Swagelok VCR metal gasket face seal fitting and two pressure gauges with different ranges.
The pump port is connected to a high flow bellow valve, which is then connected to a turbo pump system.
The bellow valve with a model number HFC2003 manufactured by Carten-Fujikon, offers a large flow coefficient of 157 and can withstand up to 25.8~bar of high pressure.
The high flow coefficient is crucial for pumping down the large vessel in a reasonable time.
The vessel's inner volume is about 600~L and is pumped with a dry fore-pump (PTS300 from Agilent) and a turbo pump (HiPace300 from Pfeiffer).
Within about 20 hours of pumping, a vacuum of less than 2E-3~Pa can be reached.
On the top flat cover, each of the seven DN-65 flanges holds one Kapton extension cable for Micromegas signals readout and for bringing the high voltage bias.
Another small DN-25 flange on the top is connected to a 1/4 inch SS pipe as gas outlet.
All flanges are designed to withstand the required vacuum and high pressure conditions.
Expanded PTFE gaskets are used for all flange fittings, including the top flat flange.

Before installing any TPC parts, extensive leak checks were carried out to verify the gas tightness of the vessel and flanges under high pressure.
15~bar of nitrogen gas with trace amount of xenon was filled into the vessel.
A Pfeiffer GSD 320 gas analysis system was used to analyze the spectrum of the air around the vessel, aiming to spot possible xenon signals.
The gas analyzer could detect tracer gas at 40 part per million (PPM) level.
No visible xenon peaks were found in the spectrum and thus no leaks were identified.
A stress test was then performed and no leaks were found during a period of 120-hour.
After installing the TPC and filling it with a working gas mixture, we kept monitoring the gas pressure over time.
No sign of leakage was found  from the vessel during the cumulated three months of operation at different pressures from 1 to 10~bar.

\subsection{Charge readout plane}
\label{sec:readout}

The charge readout plane consists of 7 Micromegas modules mounted on a circular aluminum holding plate.
The Micromegas of each module is a square piece of film with a thickness of about 0.2~mm, comprising layers of Kapton and copper.
We designed the Micromegas film and sent the drawings to CERN where fabrication was done using the Microbulk technique~\cite{Andriamonje:2010zz}.
In Section~\ref{sec:MM} we will describe Microbulk Micromegas (MM) modules with great details.
The back plate is 880~mm in diameter and 6~mm thick.
It has 7 sets of mounting and positioning holes for corresponding MM modules.
The Micromegas modules are arranged in 2-3-2 configuration, as shown in Fig.~\ref{fig:chargeReadoutPlane}, viewed from the bottom up.
The nominal gap between Micromegas modules is about 1~mm, due to machining precision of the holding plate and the modules.
By design, the flexible signal lines (shown in Fig.~\ref{fig:MMFilm}), which we call Micromegas tails, are folded to the back of the modules and go through openings on the holding plate.
MM mounting and positioning holes as well as openings for MM tails can be seen in the CAD drawing of the holding plate in Fig.~\ref{fig:fieldCage} (Left).

Fig.~\ref{fig:chargeReadoutPlane} (Left) also shows field cage copper rings, which are the boundary of the active volume.
Micromegas modules extend slightly beyond the active volume in four corners but do not cover the active volume fully in other places.
In the open areas, four rectangular open slots are added to hold photomultiplier tubes (PMTs).
Each slot is 102~mm long and 34~mm wide and can hold three 1-inch square PMTs.
PMTs can be used to measure the timing of primary ionization (usually called $t_0$) and study the absolute Z position of particle tracks.
PMTs will be mounted in future runs of the TPC.
During the commissioning runs we mentioned here, all the open slots were covered with copper foils to avoid distortions of the electric field nearby.

The holding plate also serves as a mezzanine plate between the top flat flange and the field cage.
The charge readout plane hangs on the top flat flange with 6 threaded rods.
In turn, the field cage and cathode are hanging on the holding plate with eight sets of Teflon supporting bars.
By fine tuning with the threaded rods, we can keep the charge readout plane parallel with respect to the top flat flange.
We can also move the TPC vertically to make sure the cathode mates perfectly with the high voltage feedthrough from the side port.
Nominally the holding plate is 9.6~cm below the bottom surface of the top flat flange for this purpose.

\subsection{Electric field shaping cage}
\label{sec:FieldCage}

\begin{figure}[tb]
  \begin{center}
    \includegraphics[width=\linewidth]{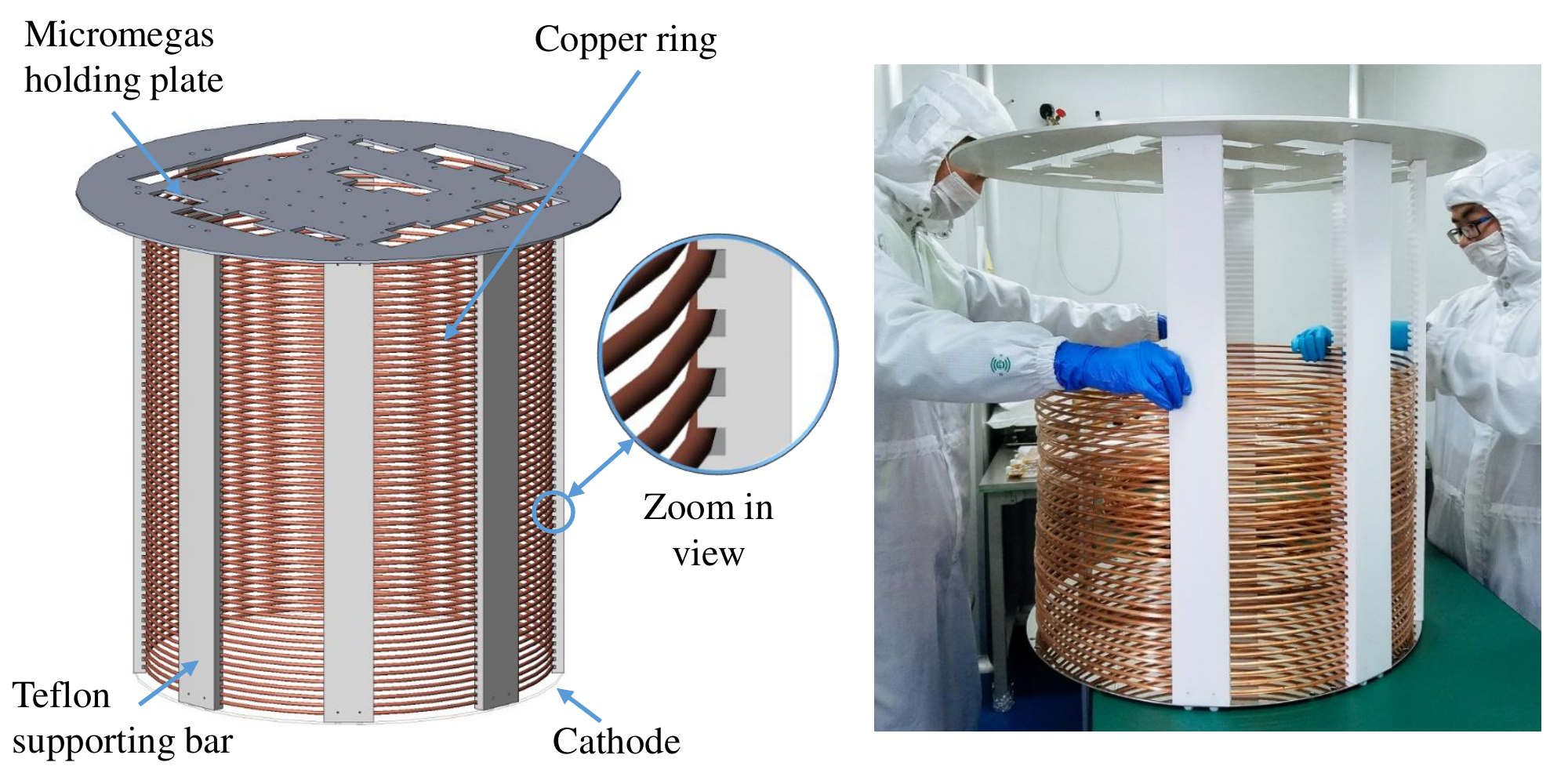}
    \caption{
   (Left): Schematic of the electric field shaping cage with main components annotated.
   It is worth noting that resistors connecting adjacent copper rings are not shown in the schematic.
   (Right): A picture of the cage during assembly. Researchers are installing copper rings with 5 out of 8 Teflon supporting bars in place.
   }
    \label{fig:fieldCage}
  \end{center}
\end{figure}

The electric field shaping cage, or field cage in short, defines the cylindrical active volume of the TPC, as shown in Fig.~\ref{fig:fieldCage}.
Copper rings serve as electrodes at graded potentials and generate an uniform field orthogonal to them.
In our TPC, electric field lines point from top to bottom and ionization electrons drift upward in the field cage.

Eight Teflon supporting bars provide the mechanical structure of the field cage.
They hang from the holding plate of the charge readout plane and the cathode is mounted to them from the bottom.
Each bar is 8~cm wide, 2~cm thick, and 78~cm long.
Their length defines the vertical height of the field cage.
The surface facing inward is grooved horizontally to hold copper rings in place.
Each groove is 6.1~mm wide and 6.1~mm deep.
The width of the ridges, or distance of adjacent grooves is 7~mm.
A zoom-in view of the grooved surface with cooper rings in place is shown in Fig.~\ref{fig:fieldCage}.

A total of 59 copper rings with a diameter of 66~cm each are biased at graded potentials.
The ring is made from copper tubes with a 6~mm diameter and a 1~mm wall thickness to reduce the weight.
The cross-section diameter is slightly smaller than the groove width on Teflon bars for easier fit.
Multiple sets of venting holes (about 2~mm in diameter) are drilled through copper tube walls to prevent air from being trapped inside.
The voltage degradation is obtained with 1~G$\Omega$ resistors placed between adjacent sets of copper rings.
Two chains of resistors are mounted for redundancy.
The overall resistance of the two resistor chains is nominally 30~G$\Omega$ and our measurements agree with the nominal value.

Our current field cage follows a widely tested design, such as in the PandaX-II dark matter experiment~\cite{Tan:2016zwf}.
In the future, we plan to test and validate alternative designs catering to the specific requirement of double beta decay searches~\cite{Chen:2016qcd, Chaiyabin:2017mis}.
The new designs would feature a solid barrel made of dielectric materials around the field cage, aiming for better insulation, and more importantly, to displace expensive enriched $^{136}$Xe gas from occupying the dead volume outside of the field cage.

\subsection{Cathode and high voltage system}
\label{sec:HV feedthrough}

\begin{figure}[tb]
  \begin{center}
    \includegraphics[width=0.8\linewidth]{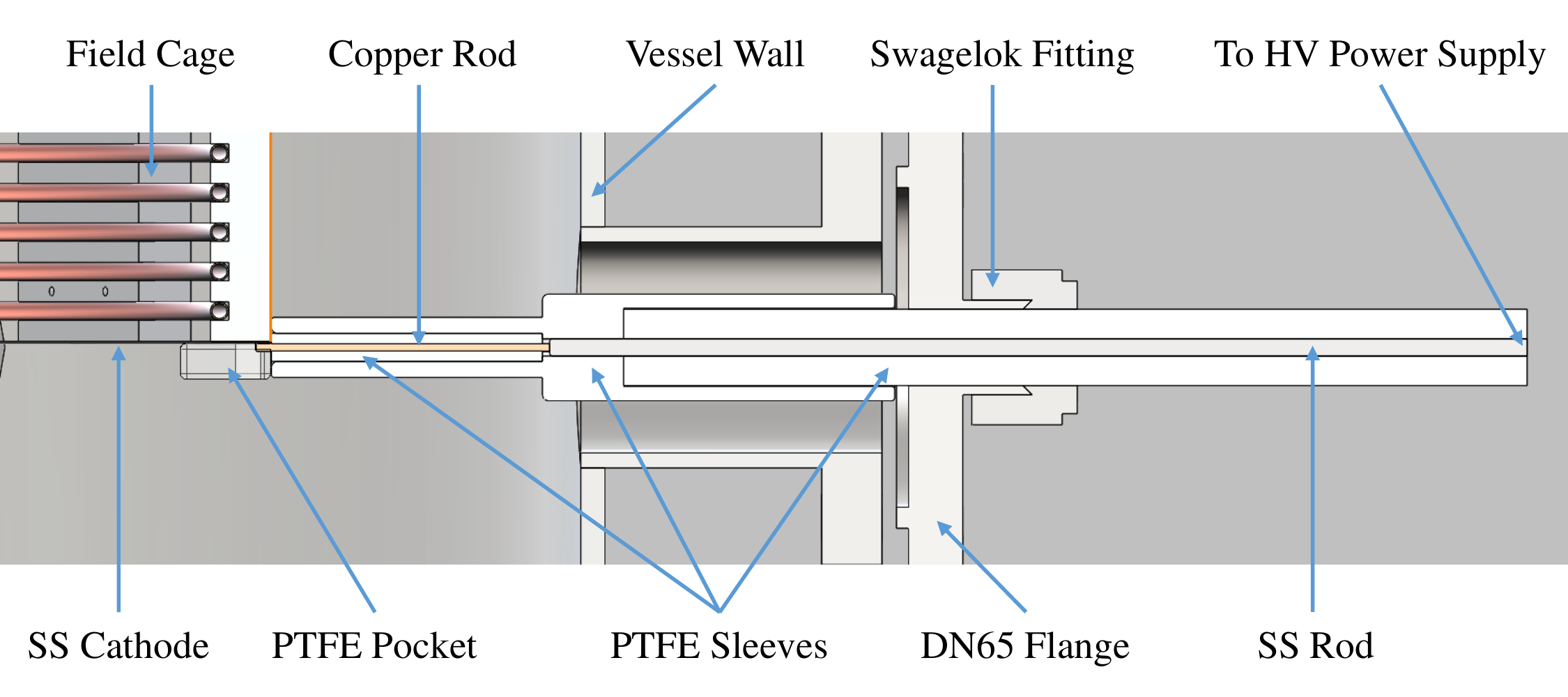}
\caption{Schematic of the feedthrough system for the TPC drift high voltage. See the main text for a detailed description of each part.}
   \label{fig:feedthrough}
 \end{center}
\end{figure}

The cathode is simply a solid stainless steel plate which is 3~mm thick and has a diameter of 694~mm.
There are 8 sets of through holes where the Teflon bars are mounted.
The plate is relatively thick and provides enough rigidity to hold the field cage in the desired cylindrical shape.
The circular edges are chamfered and all surfaces are polished to minimize electric discharge at high voltage.
Its thickness and mechanical strength also simplify all the surface treatment.

%%%%%%%%%%%%%%%%%%%%%%%%%%%%%%%%%%%%%%%%%%%%%%
% mechanical properties
%%%%%%%%%%%%%%%%%%%%%%%%%%%%%%%%%%%%%%%%%%%%%%
Negative HV on the cathode is provided through a high pressure feedthrough on one of the service ports on the vessel's side wall.
As illustrated in Fig.~\ref{fig:feedthrough}, the feedthrough is constructed using a compression seal approach.
A SS rod with a diameter of 6~mm is embedded in a Teflon sleeve, which in turn is press-fitted into a Swagelok 1~inch tube fitting.
A Swagelok male nut is welded on a blank flange.
The Teflon tube, with two Swagelok ferrules on it, is press-fitted into the nut and then clamped by a Swagelok female nut.
The assembly is then installed on a side service port.
Inside the vessel, the SS rod is connected to a copper rod, which is wrapped in PTFE sleeves for insulation purpose.
The copper rod is pressed on to the cathode plane from below by a PTFE pocket, which is used to minimize electric discharge.
This Swagelok fitting assembly has been tested extensively to withstand 10~bar pressure.

%%%%%%%%%%%%%%%%%%%%%%%%%%%%%%%%%%%%%%%%%%%%%%
% Electrical properties
%%%%%%%%%%%%%%%%%%%%%%%%%%%%%%%%%%%%%%%%%%%%%%
The high voltage for the cathode is supplied by a Matsusada AU-100N3-L HV power supply.
With the field cage length of 780~mm, a voltage of 78~kV is required to obtain an electric field of 1~kV/cm, which is our design goal.
A test in atmospheric air was performed first to optimize the HV feedthrough and coupling between feedthrough rod and cathode.
With -79.5~kV applied on the cathode in open air, a stable current leak to the ground was about 0.01~mA, which was the current measurement resolution of the DC power supply.
A similar test was performed with 10~bar nitrogen gas.
With -95.0~kV applied on the cathode, no current reading from the power supply was recorded for the whole test period of 120 hours, which means that the leakage current was less than 0.01~mA.

\subsection{Electronics and DAQ system}
\label{sec:daq system}

\begin{figure}[tb]
  \begin{center}
    \includegraphics[width=\linewidth]{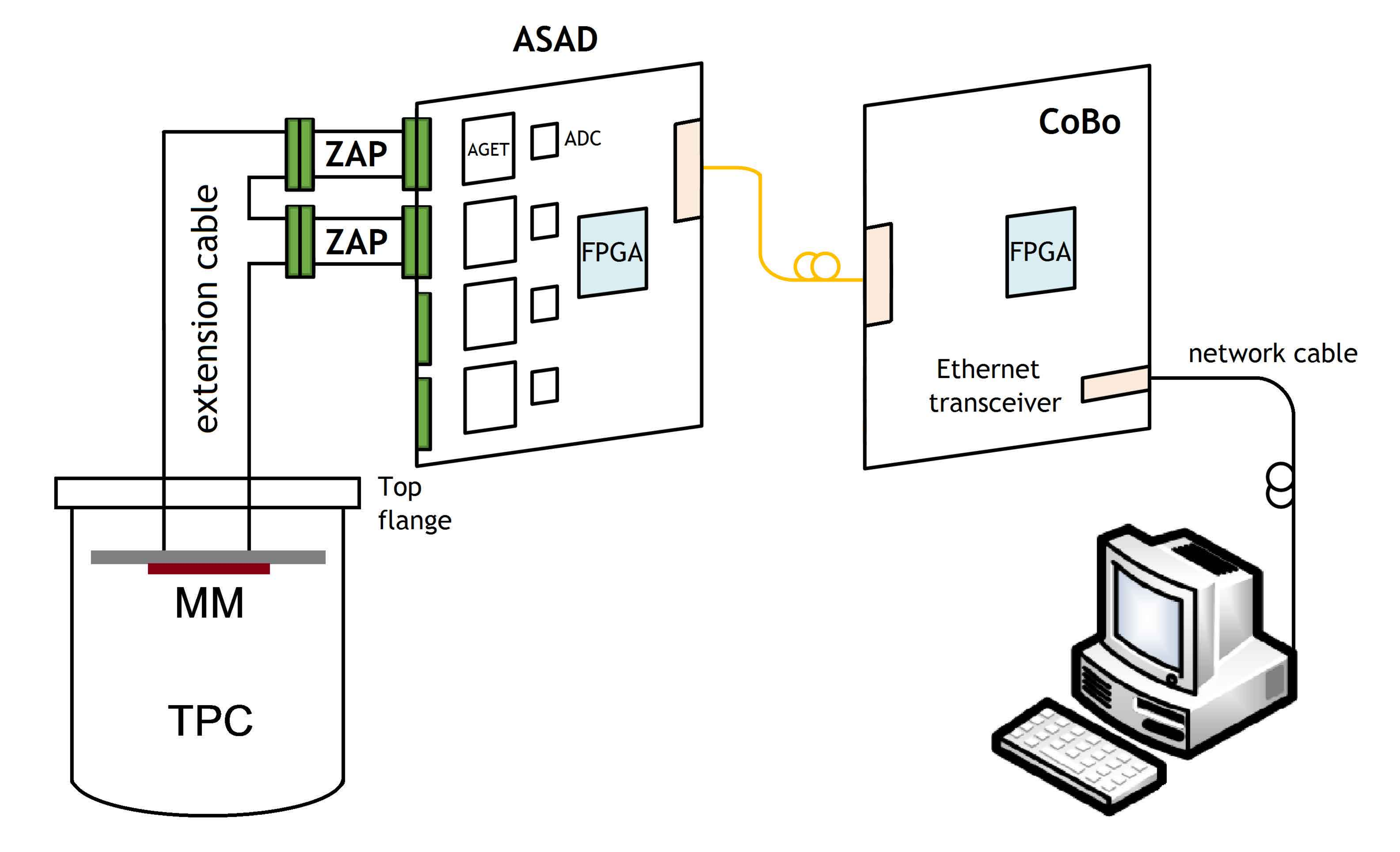}
    \caption{
   The sketch of the electronics system with AGET chips, ASAD FEC and CoBo back-end.
    }
    \label{fig:electronics}
  \end{center}
\end{figure}

The electronics and DAQ system read out signals from Micromegas when triggered, and then digitize and write data to disk.
Signals from Micromegas mesh and the readout electrode, which is segmented to 128 readout strips, are recorded.
Strip signals are digitized by a commercial ASAD/CoBo solution~\cite{Giovinazzo:2016ikh} based on AGET chips~\cite{ref:AGET_ASIC}.
An externally supplied high voltage is applied to the mesh of the Micromegas via an Ortec 142A preamplifier, where the mesh signal can also be amplified for timing and energy measurement.
During the commissioning stage of our experiment, we have not used mesh signal for triggering or spectrum analysis.

The electronics system overview is shown in Fig.~\ref{fig:electronics}.
AGET, short for ASIC for Generic Electronic system for TPCs, has wide-spread application in particle and nuclear physics experiments.
The chip can handle 64 channel input, and has a sampling frequency up to 100~MHz, a dynamic range from 120~fC to 10~pC, and peaking time from 50~ns to 1~$\mu$s.
AGET chips can tolerate a modest event rate of about 1~kHz, which is adequate for rare event searches we are interested in.
An ASAD (ASIC Support \& Analog-Digital conversion) board hosts 4 AGET chips, each of which is coupled with an external 12-bit ADC.
Therefore an ASAD board can process up to 256 Micromegas channels.
A Field-Programmable Gate Array (FPGA) on the ASAD board controls the AGETs and sends the packets of acquired data to the back-end card CoBo (Concentration Board).
A CoBo board controls and reads up to 4 ASAD boards, i.e., 1024 channels.
Besides external triggering, AGET can also compare channel inputs with preset trigger thresholds and generate the so-called \emph{multiplicity} self-trigger, where multiplicity means the number of channels over threshold.
We use self-triggering from strip signals for all our measurement.

\begin{figure}[tb]
  \begin{center}
    \includegraphics[width=\linewidth]{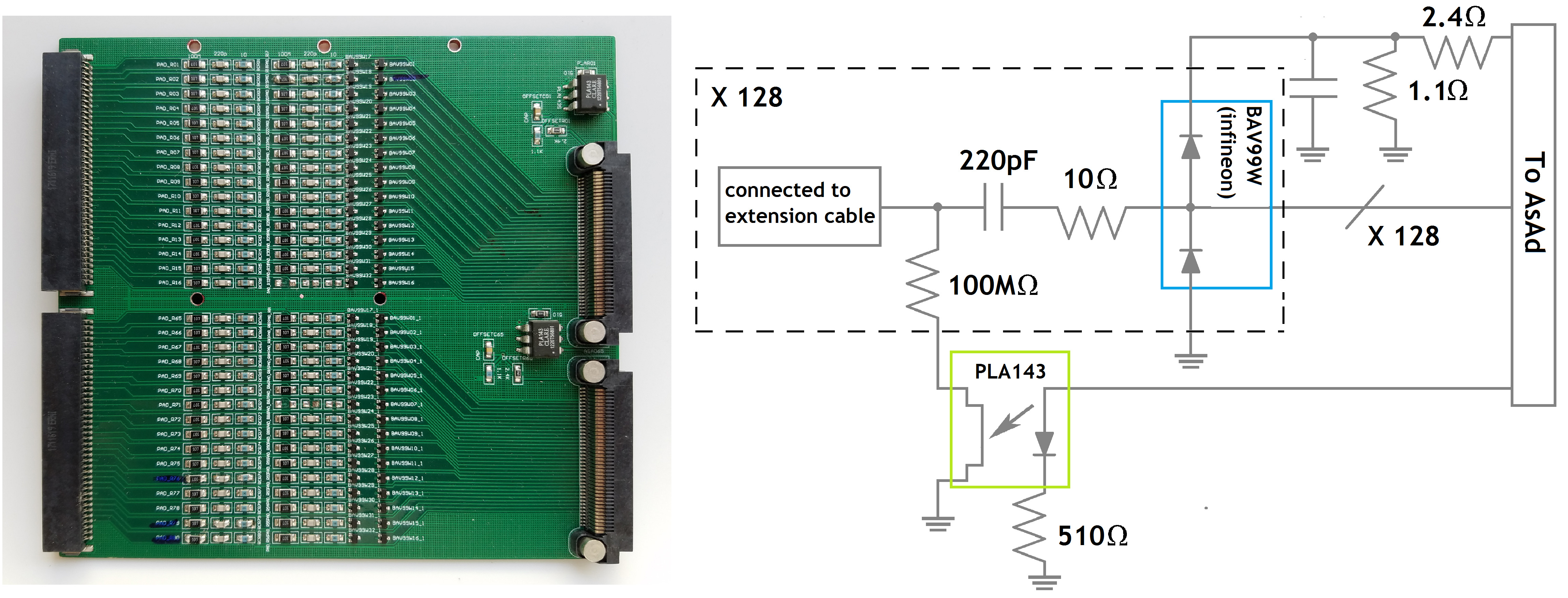}
    \caption{
   (Left): Picture of a ZAP board. On the left side, ZAP connects to a Kapton extension cable via two 80-pin 1.27 mm pitch connectors from ERNI. On the right side, ZAP connects to ASAD board via two 80-pin Samtec ERF8 connectors.
   (Right): Schematic of the circuits on the ZAP board. It follows the recommended design of our electronics vendor. Enclosed in the dotted rectangle is one sample of the 128 identical circuits. In the blue rectangle are fast switching diodes manufactured by Infineon and in the green rectangle is a relay manufactured by CLARE.
    }
    \label{fig:ZAP}
  \end{center}
\end{figure}

In our setup, we have at most 896 strip readout signals from 7 Micromegas modules.
A Micromegas module transmits signals to an ASAD board via a flexible cable and a ZAP (siZed connections And Protections) board.
The flexible cable, a polyimide-based PCB with two connectors at the ZAP end but no other electric components, extends the signal from inside the pressure vessel to outside, where it is connected to a ZAP board.
The ZAP board is a critical interconnection between the detector and ASAD since it provides the necessary protection to the ASAD against the high voltage sparks that can occur in the Micromegas.
Our ZAP board (shown in Fig.~\ref{fig:ZAP}) consists of 128 identical circuits with pull down resistors and fast switching diodes to prevent large current discharge from saturating or destroying ASAD channels.

The DAQ software we use is shipped with ASAD/CoBo and developed by NSCL (MSU) and IRFU (CEA).
It reads and writes configuration parameters of ASAD/CoBo suite, manages the data acquisition process, monitors data quality, and displays acquired events periodically.
The raw binary data produced by the ASADs are stored in separate files that are combined and converted to ROOT data format~\cite{Brun:1997pa} offline.
A graphical user interface is also provided for data display and monitoring.
We modify the software slightly to avoid event display from draining computing resources during a long run.

\subsection{Gas handling system}
\label{sec:gas system}
\begin{figure}[tb]
  \begin{center}
    \includegraphics[width=\linewidth]{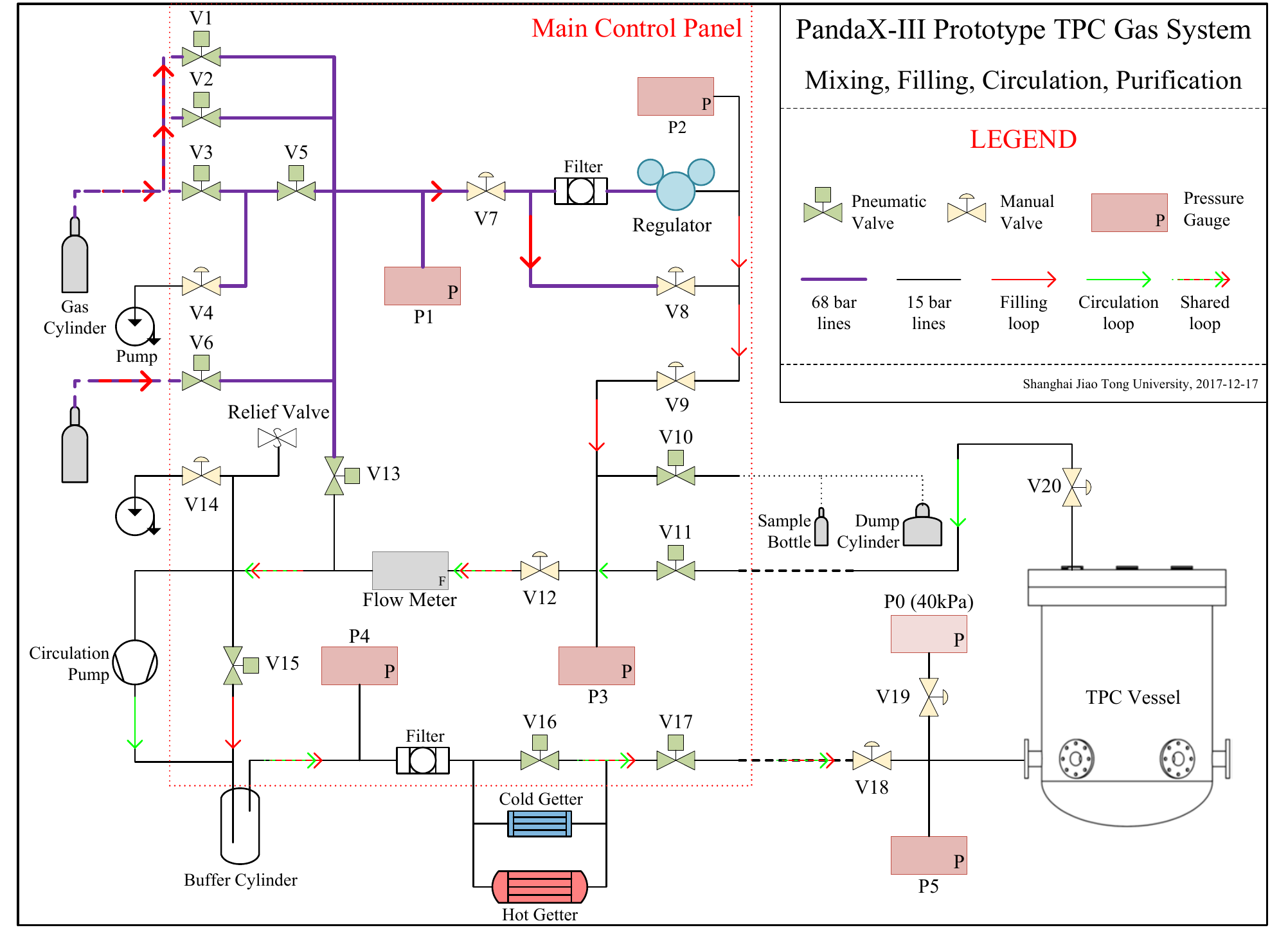}
    \caption{
   Schematic drawing for the gas handling system. Red arrows denote the filling loop. Green arrows denote the circulation loop. Dotted red and green arrows are for the shared loop. Before the regulator, all gas pipes and fittings are rated up to 68~bar.
After the regulator, the working pressure is limited to 15~bar.
    }
    \label{fig:gassystem}
  \end{center}
\end{figure}

The gas handling system performs gas filling, circulation, mixing, purification, and recovery.
All the pipes are 1/4 inch SS pipes with Swagelok VCR metal gasket face seal fittings.
All the components including valves, circulation pump, purifiers, flow meters, and gauges, have design working pressures of at least 15~bar.
The system~\cite{Chen:2016qcd} was initially designed and constructed for the PandaX-III 200~kg module.
It is well suited for the prototype TPC with minor modifications.

The schematic drawing of the system is shown in Fig.~\ref{fig:gassystem}.
Red arrows show the direction of gas flow while filling and green arrows for circulation.
Gas cylinders can be connected to the system from valves V1, V2, V3, and V6, as shown in the top left section of the figure.
Gas is filtered by a 0.5~micron particulate filter before going through a regulator, a couple of valves, and a flow meter.
During gas filling, valve V15 is open and the circulation pump is bypassed.
The 2~L buffer cylinder in the pipe line is installed to prevent sudden pressure changes.
After the buffer and another particulate filter, filled gas would go through one of two getters for purification if necessary.
The two getters, model HP190-702FV by SAES and 7NG by SIMPure, run at room temperature and high temperature respectively and can remove different kinds of impurities from gases.
The SAES room temperature getter purifies N$_2$, Ar, Xe, isobutane, TMA, SF$_6$, etc. by removing impurities such as H$_2$O, O$_2$, CO, CO$_2$, and H$_2$.
Besides the above mentioned impurities, The high temperature SIMPure getter removes N$_2$ and hydrocarbon compounds (including isobutane and TMA).

The lower half of the pipe line is also used as part of the circulation loop.
Due to outgassing, the quality of the gas mixture in the TPC vessel deteriorates over time.
To mitigate this, continuous circulation and purification are often desirable.
For this operation, V9 has to be closed to isolate upstream pipes.
V20 and V11 have to be open and the circulation pump pushes gas mixture to circulate.

Gas recovery can be performed in the same line for gas filling.
A large 220~L stainless steel gas cylinder can be connected to the system via valve V10 for recovery of precious gas, such as xenon.
When immersed in a liquid nitrogen dewar during operation, the cylinder functions as an absorption pump and liquifies xenon for storage.

When preparing the Ar-(5\%)isobutane mixture for our commissioning runs, the two kinds of gas were mixed directly in the vessel.
Firstly we filled 5~kPa isobutane gas into the TPC vessel.
The pressure was measured with a high precision pressure gauge mounted on one of the side service ports on the vessel.
The gas system was pumped down again to vacuum to get rid of the residual isobutane in the pipes.
Finally we filled the vessel with argon gas to 1 bar.

Xe-(1\%)TMA mixture was mixed differently because xenon needs purification by going through the high temperature getter first.
Five~bar xenon gas was filled into the TPC vessel first and purified by the high temperature getter for about a week of continuous circulation, to remove the impurities from xenon gas itself.
Afterwards the purification process was stopped and an empty sample bottle was connected to V10 for temporary TMA storage in next step.
A large cylinder with pure TMA was then connected to the sample bottle.
The sample bottle was cooled down with liquid nitrogen and the desired amount of TMA, determined by weight, was allowed to liquify inside the sample bottle.
We then isolated the sample bottle from the TMA cylinder and opened V10 to allow TMA to diffuse to the gas circulation system.
The TMA sampling bottle was heated to speed up the diffusion process.
Meanwhile the purification process restarted with the room temperature getter to purify the mixture, because the high temperature one could remove TMA also.
The purification process continued through all the commissioning runs with Xe-(1\%)TMA mixture.

Future upgrades are planned for the gas system.
We will add a smaller 20~L scale recycling gas cylinder for an easier operation.
Precision gas flow meters will be installed for another way of measuring gas mixture ratio.
We will also design and implement an online gas composition analysis system with a mass spectrometer for yet another measurement of gas content.

\section{Microbulk Micromegas Modules}
\label{sec:MM}

\begin{figure}[tb]
  \begin{center}
    \includegraphics[height=0.55\linewidth]{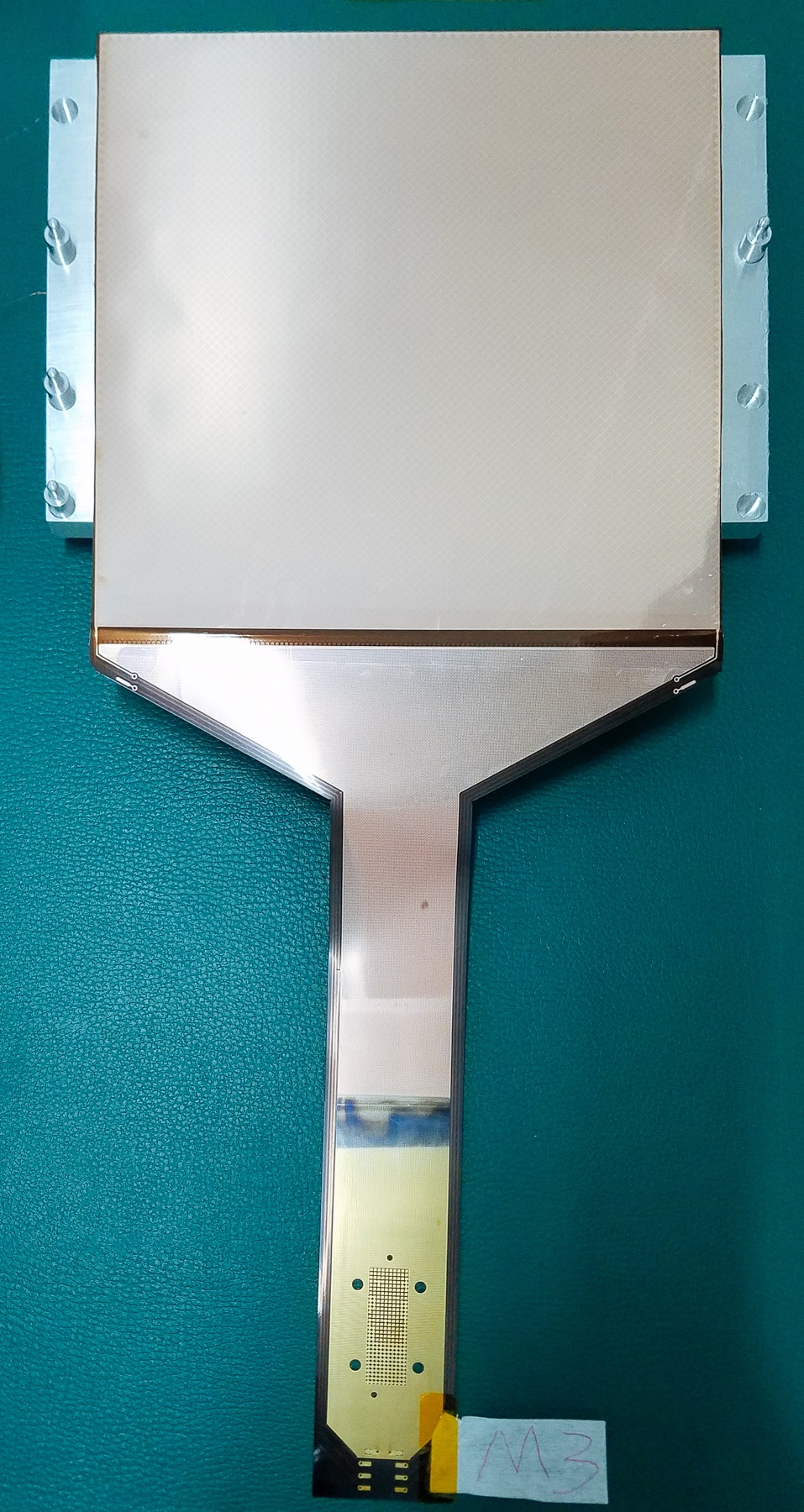}
    \hspace{3em}
     \includegraphics[height=0.55\linewidth]{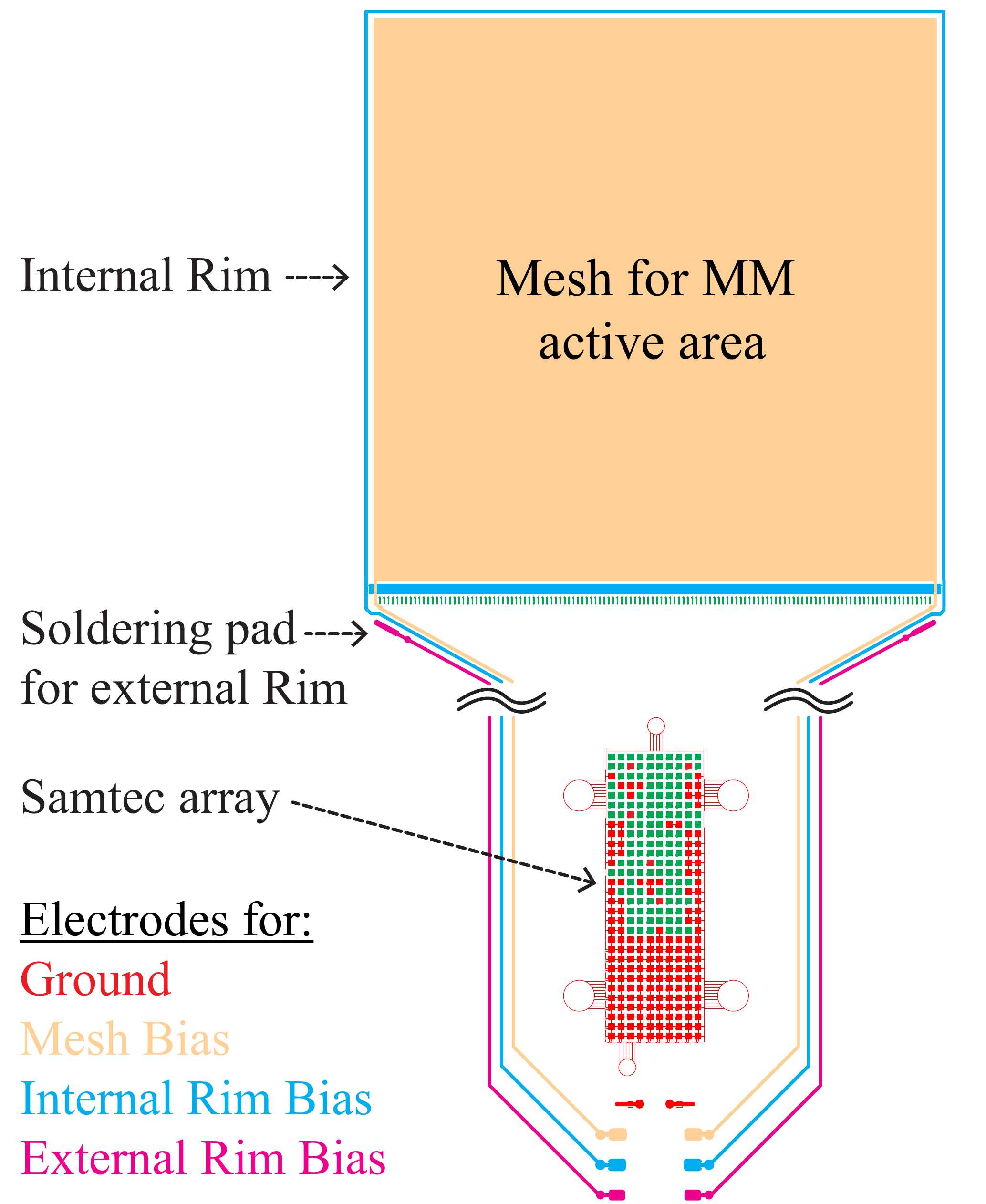}
    \caption{
   (Left): Picture of a Micromegas detector fabricated with Microbulk technique.
   The square active area is 20 cm by 20 cm.
   Zoom-in view and illustration of the active area are shown in Fig.~\ref{fig:MMStrips}.
   (Right): An illustration of bias and readout electrodes for the Micromegas film.
   The tail and the MM active area are not to scale.
   Green pads in the Samtec high density solderless interface array footprint are for channel readouts.
   The array has 300 pins and we are utilizing 128 of them.
   Four sets of electrodes for ground, mesh bias, internal rim bias, and external rim bias respectively are color-coded and annotated in the sketch.
   See the main text for detailed descriptions of the electrodes.
    }
    \label{fig:MMFilm}
  \end{center}

\end{figure}

\begin{figure}[tb]
  \begin{center}
    \includegraphics[width=0.4\linewidth]{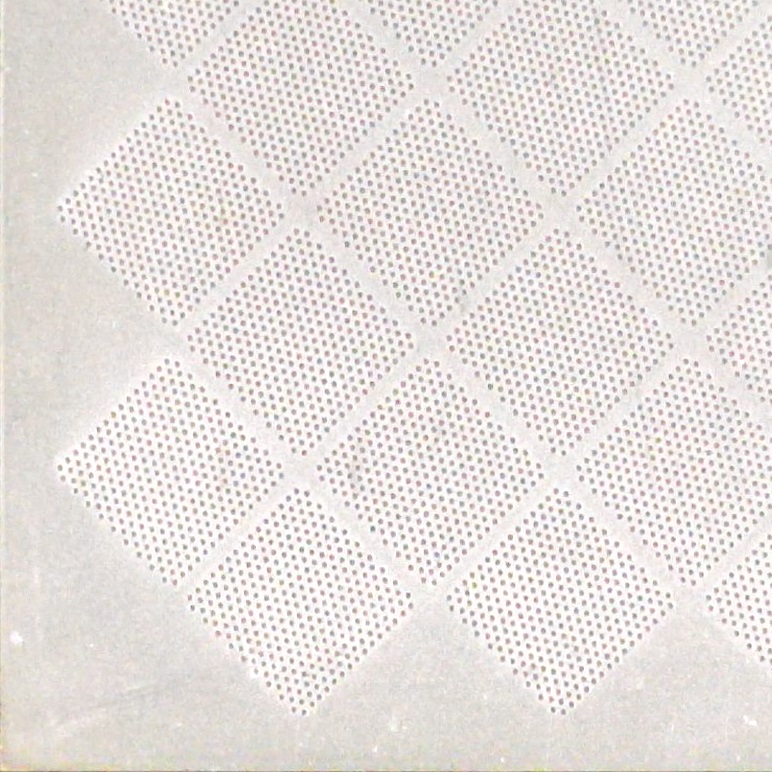}
    \hspace{1em}
    \includegraphics[width=0.45\linewidth]{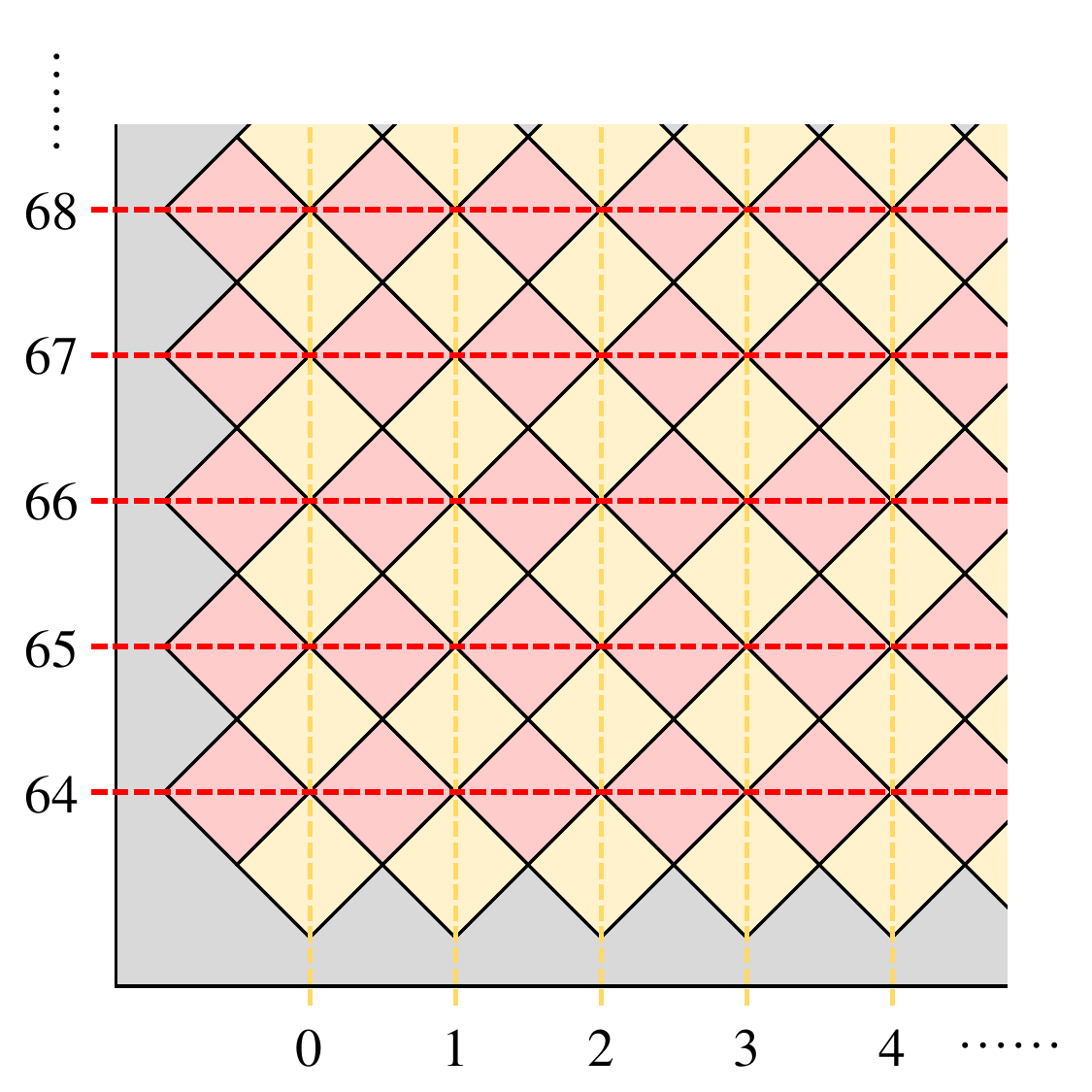}
    \caption{
   (Left): Zoom-in view of Micromegas mesh at one of the corners of the active area. Amplification holes are clustered in diamond-shaped pixels, whose diagonals are 3~mm.
   (Right): An illustration of Micromegas strip readout scheme under the mesh. Yellow vertical lines and red horizontal lines are drawn to show X and Y strips connecting the corresponding pixels on mesh.
    }
    \label{fig:MMStrips}
  \end{center}
\end{figure}

Micromegas is widely used in gas detectors to amplify and readout electron signals~\cite{Giomataris:1995fq}.
Recently, the Microbulk fabrication technique~\cite{Andriamonje:2010zz}, which utilizes standard printed circuit lithography work-flow, has been developed to make Micromegas with layers of Kapton and copper.
For MM, mesh and electrode planes are separated not by commonly-used pillars, but by a Kapton film.
The film is etched to form many groups of small holes where the avalanches happen.
The height of the amplification gap of MM is determined by the thickness of the Kapton film and its uniformity over the whole active area can be well controlled.
For this reason, MM has excellent intrinsic energy resolution compared to traditional fabrication techniques.
The radioactive contamination, which is a critical index for low-background rare-event-search experiments, can also be lower thanks to material selection and small mass to active area ratio.

The Microbulk Micromegas films we use are produced by the Micro-Pattern Technologies (MPT) group at CERN.
The amplification gap is 50~$\mu$m.
The active area is about 20~cm by 20~cm, as shown in the picture in Fig.~\ref{fig:MMFilm}.
What is visible in the picture is the mesh of MM.
If we zoom in at one corner, one can see that the mesh has a large number of tiny holes (50~$\mu$m in diameter) clustered in groups (Fig.~\ref{fig:MMStrips} (Left)), and each group corresponds to one \emph{pixel} of the Micromegas.
The underlying electrode is segmented into intervening strips.
Pixels along X or Y direction are grouped together and read out as a strip (Fig.~\ref{fig:MMStrips} (Right)).
The pitch of a pixel, defined as the diagonal distance of the diamond-shaped footprint, is 3~mm.
Overall we have 128 strip channels where half are in X direction and half in Y direction.
It is worth noting that energy deposition on the MM is shared between X and Y strips.

All strips are read out from a long tail through a Samtec high density solderless interface array (see Fig.~\ref{fig:MMFilm}).
The Samtec connects two mirror-imaged pad arrays with spring-loaded contacts when pressed together.
On the very end of the MM tail, there are four pairs of electrodes for ground, mesh high voltage bias, internal rim bias, and external rim bias (Fig.~\ref{fig:MMFilm} (Right)).
The external rim is embedded in the back-plate underneath the MM film and circumvents three edges that have no tails.
The internal rim is fabricated along with the Micromegas film itself and circumvents all of the four edges of the active area.
During normal operation, readout strips on the anode are connected to the ground with protecting resistors in between and the mesh is biased with a negative high voltage of a few hundred volts.
By applying a slightly higher (negative) voltage to the rims, they would shape electric field lines away from the edge of the MM and thus prevent most of the drift electrons from arriving at the dead area.

Micromegas can be tiled together to cover a larger area, as what we are doing in our readout plane (see Fig~\ref{fig:chargeReadoutPlane}).
The tails of MM have been folded to the back of the active area to avoid interfering with nearby modules.
%The space between two Micromegas modules is nominally about 1~mm and would be the so-called \emph{dead-zone}.
%However, the dead-zone effect can be mitigated with the aforementioned rims.

\subsection{Micromegas module gluing}

\begin{figure}[tb]
  \begin{center}
    \includegraphics[width=0.65\linewidth]{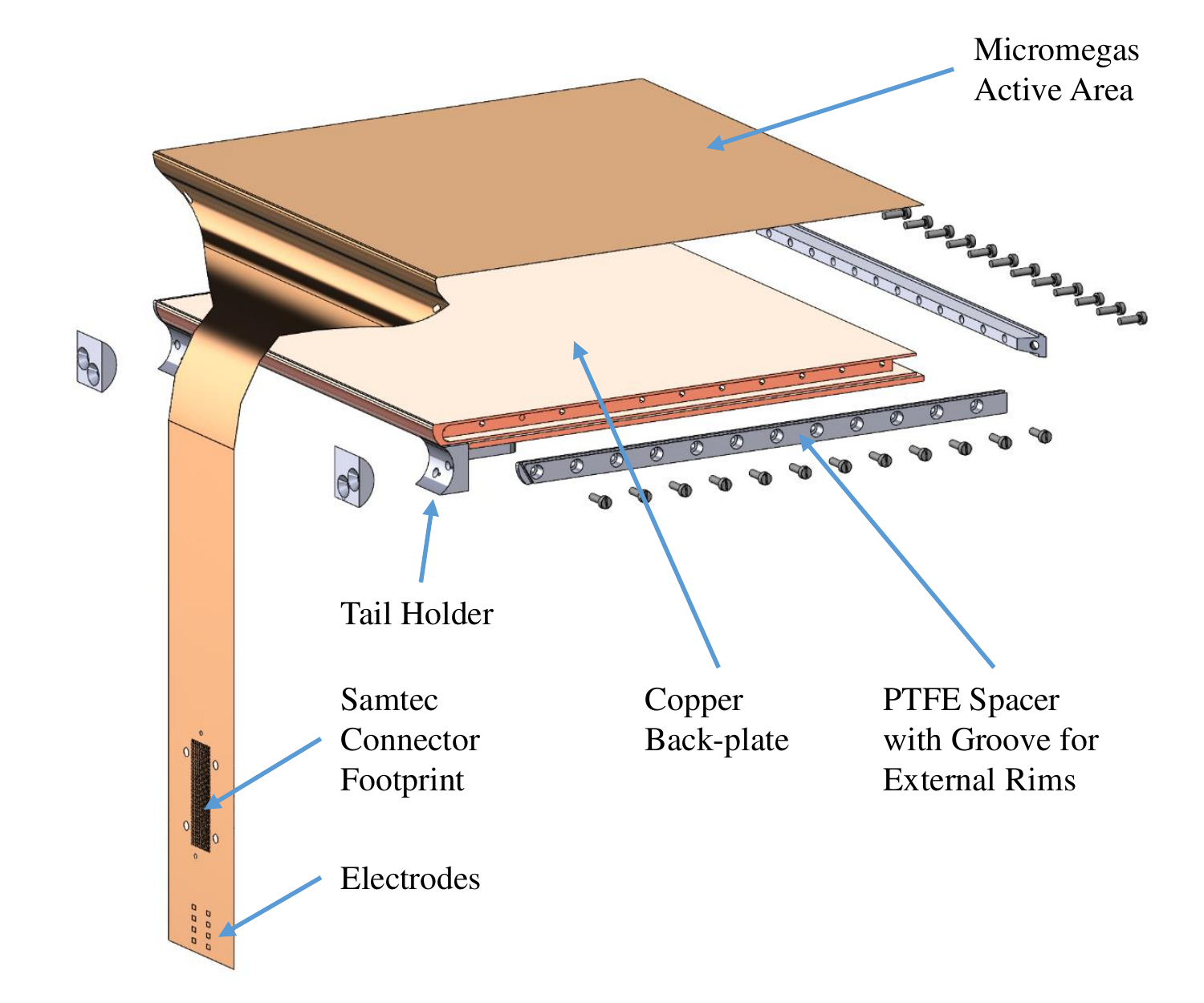}

    \caption{
   An exploded view of a Microbulk Micromegas film on the copper back-plate.
   The copper plate has PTFE spacers with thin grooves to accommodate an external rim.
   Signals from 128 strip channels are connected to a 300-pin Samtec connector (GFZ series).
    }
    \label{fig:SR2MExplodedView}
  \end{center}

\end{figure}

Microbulk Micromegas is built with only micron-thin layers of Kapton polyimide and copper, so its flexibility is similar to commonly seen flexible printed circuit boards.
The Micromegas films should be held rigidly perpendicular to the drift electric field and free from any surface bumps.
To do so we first glued each Micromegas film on a back-plate and then we mounted each assembly on the frame of the charge readout plane.
The back-plate is 1~cm thick and has a flat surface of 20~cm by 20~cm, which matches the footprint of Micromegas active area exactly.
An exploded view of the Micromegas module is shown in Fig.~\ref{fig:SR2MExplodedView}.
The bulk of the back-plate is made of copper with Teflon strips embedded in three edges.
Along the edges a copper wire is buried and serves as an external field shaping rim to further push drifting electrons away from possible gaps between different Micromegas modules.

The gluing was done on an auxiliary stand with positioning pins for better alignment (see Fig.~\ref{fig:MMFilm}).
Correspondingly there are positioning holes on the Micromegas film outside the active area.
In the final process the extensive area of the Micromegas film was cut off, leaving the active part matching the back-plate.

\subsection {Kapton extension cable}

\begin{figure}[tb]
  \begin{center}
    \includegraphics[height=0.25\linewidth]{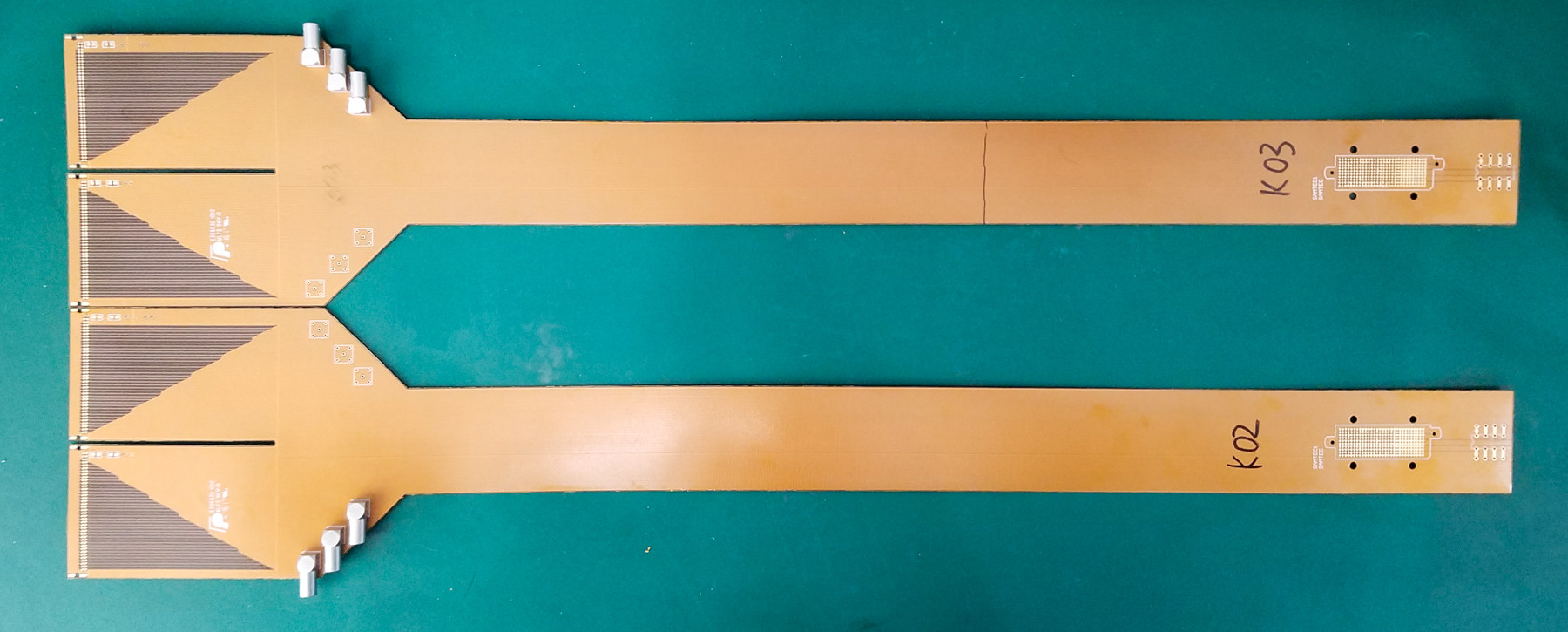}
    \includegraphics[height=0.25\linewidth]{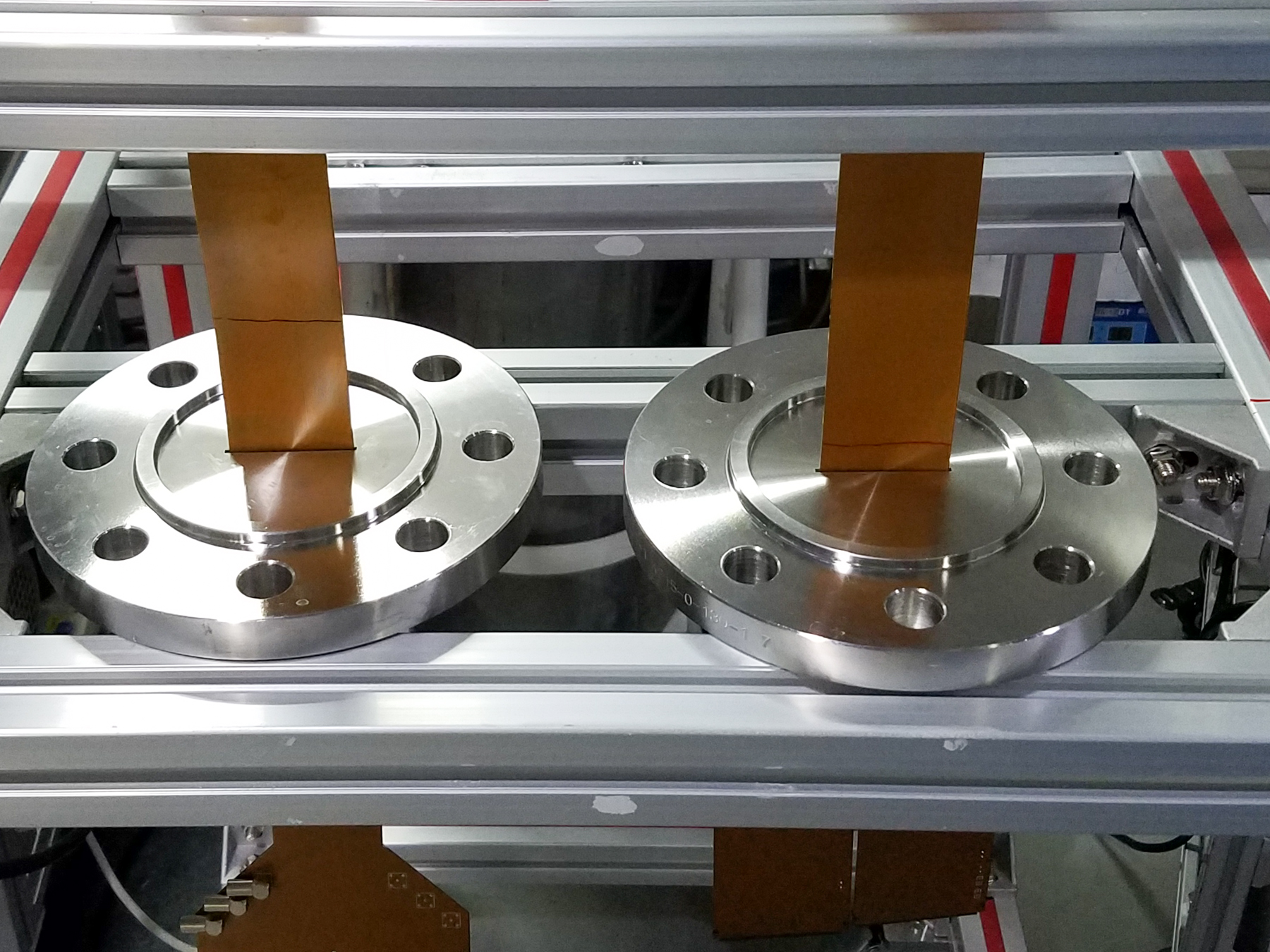}
\caption{(Left): Picture of the Kapton extension cables.
(Right): Picture of the signal feedthrough, which is made of an extension cable glued in a flange.}
   \label{fig:KaptonCable}
 \end{center}
\end{figure}

\begin{figure}[tb]
  \begin{center}
   \includegraphics[width=0.99\linewidth]{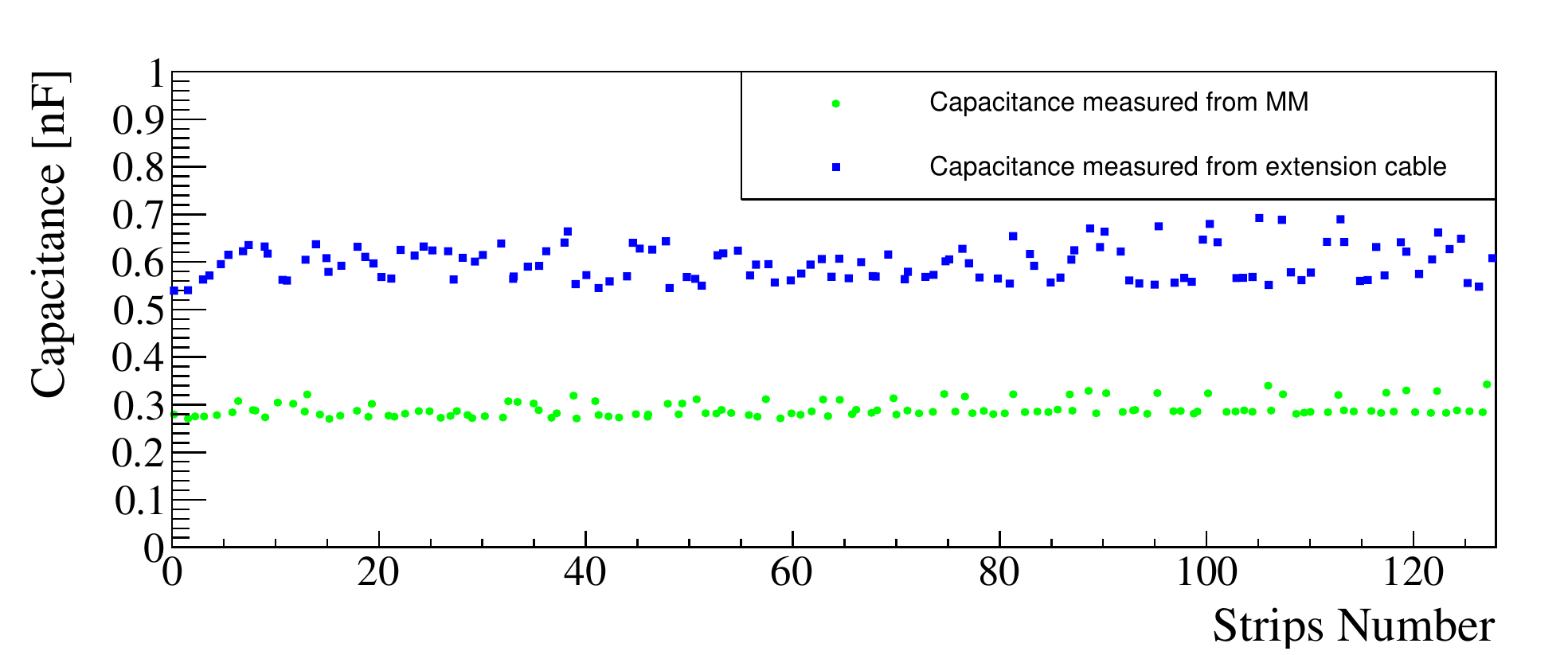}
    \caption{
   (Top): Measured capacitance between each strip of the Micromegas and ground.
   (Bottom): Measured capacitance from the extension cable.
   All capacitance values fall within the expected range.
    }
    \label{fig:connection}
  \end{center}
\end{figure}

The Kapton extension cable extracts signals from the Micromegas to the front end electronics and applies the externally supplied high voltage to the Micromegas mesh and rim electrodes.
It is also the center piece for the feedthrough between high pressure gas mixture inside the pressured vessel and the outside atmosphere.
An extension cable is necessary for three practical reasons.
Firstly, the length of the MM signal tail is limited by the fabrication process and is not long enough to reach the outside of the vessel easily.
Secondly, the assembly procedure with an extension cable in the feedthrough is much more straightforward than the case with MM signal tail in the feedthrough.
Thirdly, the extension cable serves as an adapter board between the Micromegas and the ZAP boards.
The extension cable brings additional capacitance and potentially higher noise contribution.

The extension cable is a Kapton-based flexible PCB.
As shown in Fig.~\ref{fig:KaptonCable} (Left), there are footprints for two ERNI double-row 80-pin board-to-board receptacle connectors (ERNI Manufacturer number 154744) on the left-hand side and a footprint for one 300-pin Samtec solderless interface array on the right-hand side.
The cable is 625~mm long and 0.6~mm thick.
The widest part on the left is 115~mm and the main body is 45~mm wide.
Signal and electrode traces are arranged in two layers, and each is sandwiched between two ground layers to protect it from noise and crosstalk.
Traces of all twelve Kapton cables produced by our vendor have been tested one by one to ensure connectivity and verify the absence of any short circuit between adjacent traces.
All high voltage lines have also been tested to withstand at least 700~V.
We had difficulties to solder the ERNI receptacle connectors on the extension cables, and some of them did not pass the connectivity test.
We are designing a newer version of the extension cable with a combination of rigid and flexible PCBs.

The extension cable is then glued into a rectangular slot on a stainless steel flange to make a feedthrough (see Fig.~\ref{fig:KaptonCable} (Right)).
The slot is 50~mm long and 2~mm wide.
After placing a cable in the middle of the slot, we poured in Stycast RE2039 epoxy manufactured by Loctite.
From the bottom, Teflon tapes were used to stop epoxy from leaking out.
The epoxy was then left curing for 24 hours at room temperature.
A couple of test feedthroughs were made and tested on a dedicated setup.
In this setup, the feedthrough was tested with 10~bar argon on one side and vacuum on the other side.
We observed the vacuum level rising over time, which was a combined effect of outgassing, leaking from various ports, and leaking from the feedthrough.
Measurements were done from 10 hours to as long as 72 hours periods.
From these tests, we measured an upper leak rate of 1.9~gram of xenon per feedthrough per year.
The whole TPC setup, with 7 such feedthroughs, has been tested under 10~bar argon or xenon gas for weeks, and we have not observed any measurable pressure change in the vessel, confirming the feedthrough's gas tightness at 10~bar.

We measured strip-to-ground capacitance channel by channel from a MM module's signal tail as well as from the extension cable to check the connectivity.
Results of such measurements for one Micromegas module are showed in Figure~\ref{fig:connection}.
Measurement from the MM's signal tail shows capacitance around 0.3~nF for all 128 channels.
Then the same measurement was performed from the extremity of the extension cable after the module had been installed in the TPC.
In this case, all capacitance values are around 0.6~nF.
Should a MM channel fail or the connection between MM and the extension cable break, the capacitance would deviate from the nominal values.
Therefore we routinely measure MM channels' strip-to-ground capacitance from the extension cable to monitor the health of MM and connectivity.

\section{Commissioning Runs with $^{241}$Am Source}
\label{sec:data-taking}

\begin{figure}[tb]
  \begin{center}
    \includegraphics[width=\linewidth]{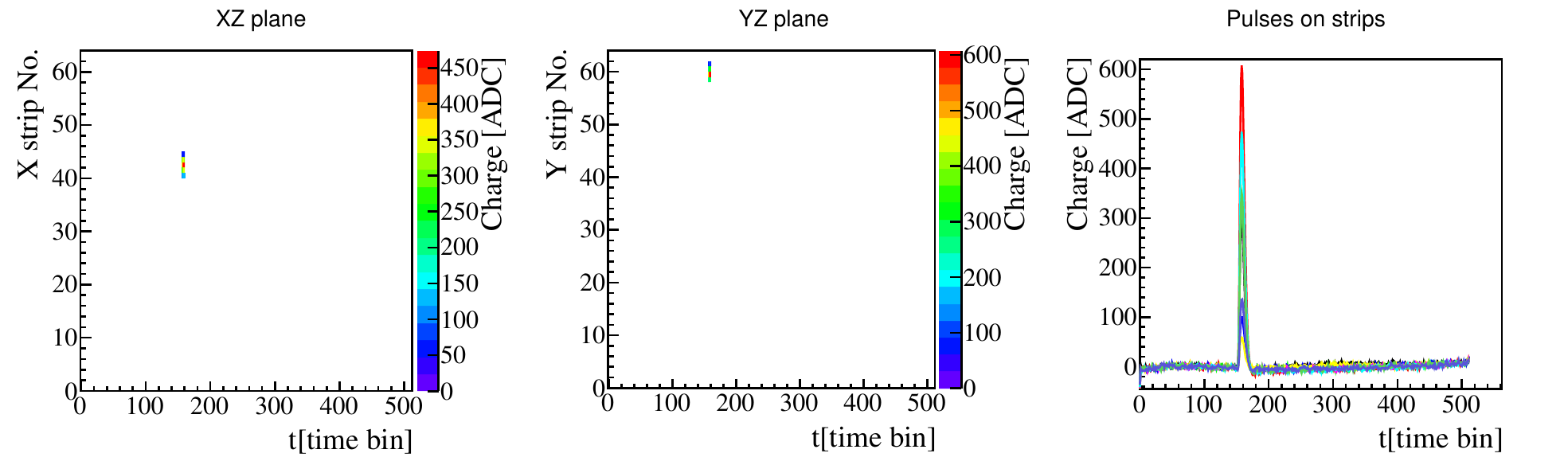}

    \caption{
    A gamma event from $^{241}$Am in 1~bar Ar-(5\%)isobutane gas mixture: The projected gamma tracks on XZ plane (Left) and YZ plane (Middle) are shown. Pulses collected on each triggered strip are shown in the right panel, where each color indicates one channel.
     }
    \label{fig:gammaSignal}
  \end{center}
\end{figure}

\begin{table}
\begin{center}
\begin{tabular}{|c|c|c|c|c|}
\hline
Run Number  & E$_{Drift}$[kV/cm]& Number of Events\\
%Type & Energy[keV]&number/decay[\%]& $\lambda_{\gamma,Ar}$(1bar)[cm]&$R_{e,Ar(CSDA)}$(1bar)[cm]\\
\hline
1 & 0.036  & 1.16$\times$10$^{6}$ \\
2 &  0.049   & 1.34$\times$10$^{6}$\\
3  & 0.062   & 1.01$\times$10$^{6}$\\
4 &  0.074  & 1.05$\times$10$^{6}$\\
5 &  0.085   & 3.41$\times$10$^{6}$\\
6 &  0.097  & 1.54$\times$10$^{6}$\\
7 &  0.110   & 1.22$\times$10$^{6}$\\
8 &  0.123   & 3.88$\times$10$^{6}$\\
9 &  0.136   & 1.49$\times$10$^{6}$\\
10 & 0.149   & 0.23$\times$10$^{6}$\\
\hline
\end{tabular}
\end{center}
\caption{\label{data-statistic} Summary of data taking with 1 bar Ar-(5\%)isobutane mixture for different drift HV configurations. The amplification field was fixed at 74~kV/cm. Run 8 with the most statistics was used to calibrated the energy resolution of MM. Its electron transmission had reached the optimal value (see description in Section~\ref{sec:electron-transmission})}.
\end{table}

We have conducted a series of commissioning runs with a $^{241}$Am calibration source, using different gas mediums at different working pressures.
Gases, such as pure argon, Ar-(5\%)isobutane, pure xenon, and Xe-(1\%)TMA, have been tested at 1~bar, 1.5~bar, 4~bar and 5~bar.
Here we report exemplary results from data taking with 1~bar Ar-(5\%)isobutane and 5~bar Xe-(1\%)TMA mixtures.

With these commissioning runs, we have been able to validate the integrality of all the sub-systems described earlier.
Key features of a TPC, such as the tracking capability, have been demonstrated.
We also made preliminary studies on the noise level of the detector, electron transmission efficiency in the drift region, and detector energy resolutions under different run conditions.

\subsection{$^{241}$Am source}
\label{sec:am-source}

A 1.43$\times10^3$~Bq $^{241}$Am source was used for detector calibration.
The source has a circular active area of 4~mm in diameter.
It was covered with a piece of 0.2~mm thick plastic film, and therefore alpha and electron emissions were shielded and it served as a low energy gamma source.
The source emits gamma rays at 59.5~keV (intensity = 35.9\%) and 26.3~keV (intensity = 2.3\%)~\cite{NNDC,LNE-LNHB}.
In addition, X-rays at lower energies such as 13.9~keV, 16.8~keV, 17.8~keV and 20.6~keV are emitted by its daughter atom $^{237}$Np.
The $^{241}$Am source was placed on the cathode plane in the Ar-(5\%)isobutane run, while it was hanged about 40~cm below the readout plane in the Xe-(1\%)TMA run.
The source was positioned intentionally at places whose projections on the readout plane avoided the broken strips of the Micromegas.

\begin{figure}[tb]
  \begin{center}
   \includegraphics[width=0.48\linewidth]{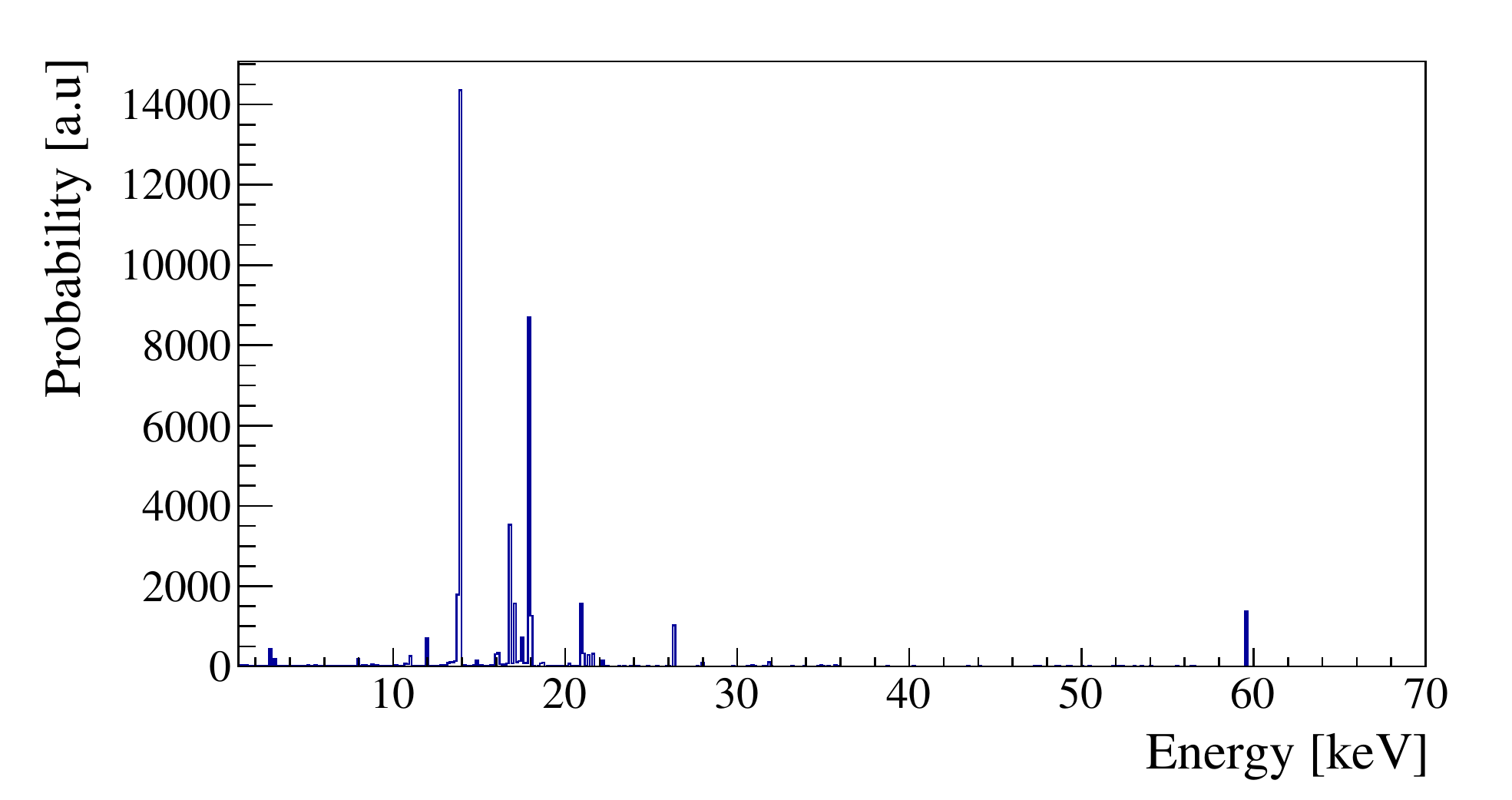}
   \includegraphics[width=0.48\linewidth]{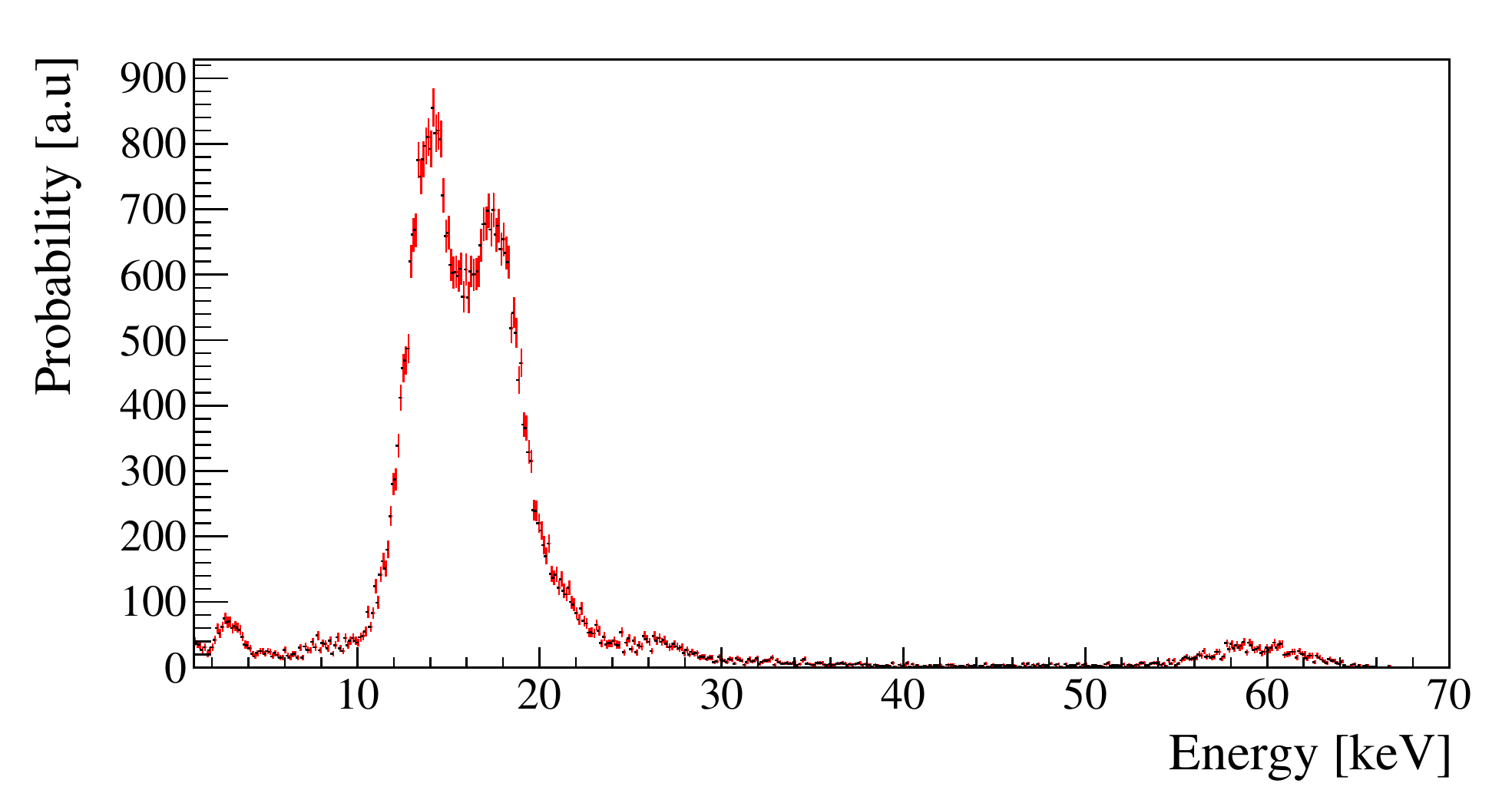}
   \includegraphics[width=0.48\linewidth]{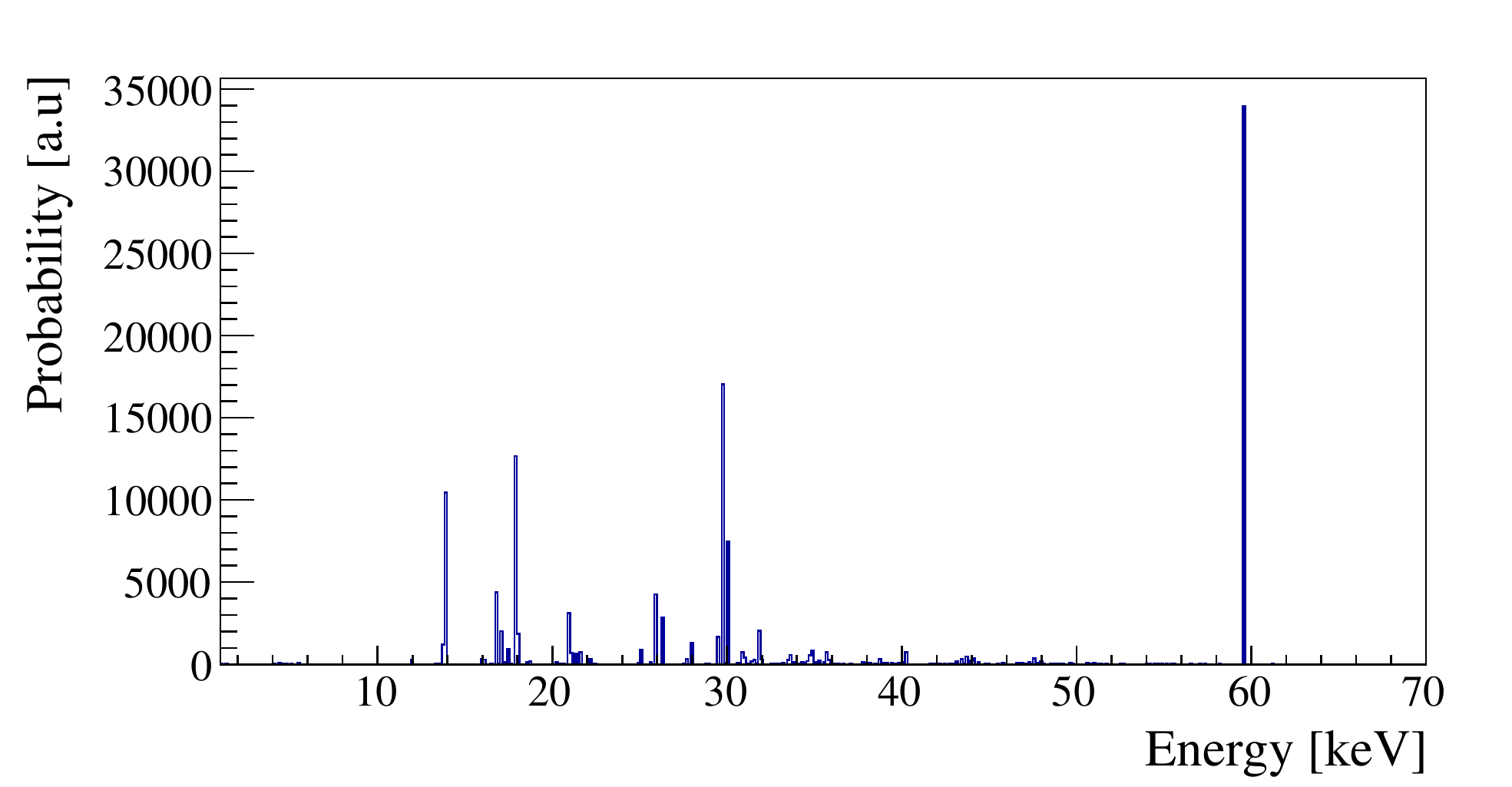}
   \includegraphics[width=0.48\linewidth]{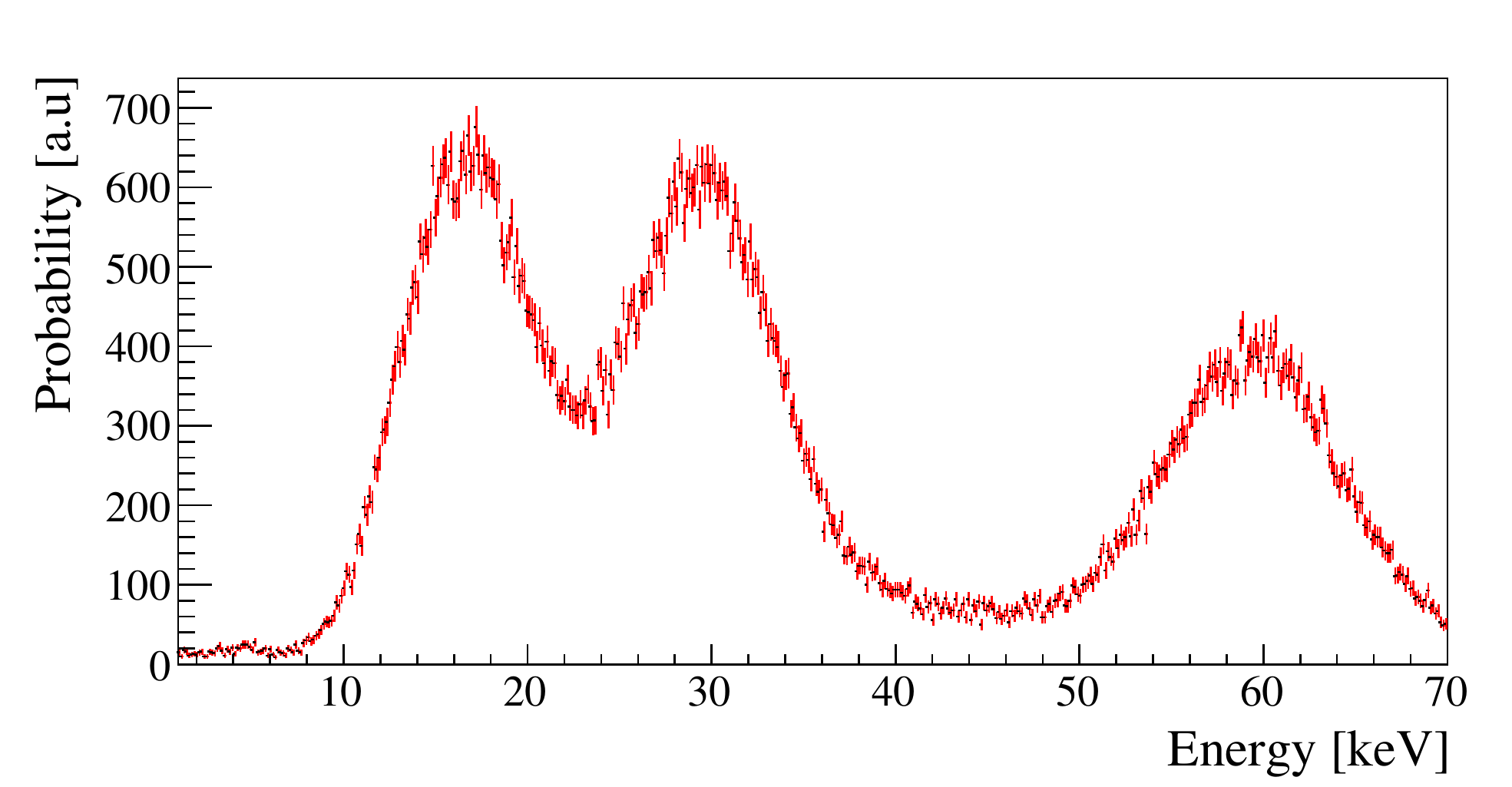}
   \caption{
   (Top): Energy spectrum lines of the $^{241}$Am source detected by one Micromegas in 1 bar Ar-(5\%)isobutane by GEANT4 simulation before (Left) and after energy smearing (Right).
   (Bottom): The same simulation but for 5 bar Xe-(1\%)TMA.
   The simulated spectra for the Xe-(1\%)TMA mixture follow the same track position cut and triggered channel number cut we used for data shown in our measured energy spectrum.
   To mimic our detector response, the energy of each event from GEANT4 is smeared randomly by a Gaussian noise term, and the sigma of the Gaussian function is proportional to $\sqrt E$.
    }
    \label{fig:simulation}
  \end{center}
\end{figure}

A GEANT4~\cite{Agostinelli:2002hh} based Monte Carlo simulation has been developed to study the expected spectrum of the $^{241}$Am source.
The simulation geometry faithfully reconstructs the main features of the TPC active volume and Micromegas arrangements on the readout plane.
Other main components such as the SS vessel, are included but with less detail.
The charge readout plane has 7 Micromegas modules in the simulation, but only one is turned on as sensitive surface, as it was in the real operation scenario.
The simulated energy spectrum in 1~bar Ar-(5\%)isobutane is shown in Fig.~\ref{fig:simulation} (Top).
The two $\gamma$-emission peaks at 26.3~keV and 59.5~keV are small and X-ray emissions of $^{237}$Np show prominent peaks in the energy range between 13.9~keV and 20.6~keV.
In the simulation for 5~bar Xe-(1\%)TMA mixture (Fig.~\ref{fig:simulation} (Bottom)), where the attenuation lengths of gamma particles and electrons are shorter, the 59.5~keV gamma peak from $^{241}$Am and the 29.0~keV Xe K$_{\alpha}$ escape peak dominate.
All other peaks in the Ar-(5\%)isobutane simulation also show up, but with different intensities.

\subsection{Data with 1~bar Ar-(5\%)isobutane mixture}
\label{sec:electron-transmission}

Initial data taking was devised with 1 bar Ar-(5\%)isobutane mixture, which is a classical gas mixture for TPC characterization and widely used in several experiments~\cite{Dafni:2009jb, Iguaz:2012ur}.
The operating HV for the Microbulk Micromegas module was $-$370~V, which corresponds to an amplification field of 74~kV/cm.
We ran at different drift fields in the stable running conditions, from 36 to 149~V/cm, to study its effect on detector energy resolution and electron transmission, as summarized in Table~\ref{data-statistic}.
A total of $1.6\times$10$^{7}$ events were recorded during the drift voltage scan.

A typical event of a low energy gamma is shown in Fig.~\ref{fig:gammaSignal}.
The AGET chip compares the input of each channel to a pre-defined threshold.
Once signals from multiple channels are over threshold, an \emph{event} is registered.
Our DAQ system records pulses of all over-threshold channels, as shown in the right panel of Fig.~\ref{fig:gammaSignal}.
In our commissioning runs, we used a sampling frequency of 5~MHz and record a length of 512 samples, resulting in a time window of 102.4~$\mu$s.

\begin{figure}[tb]
  \begin{center}
    \includegraphics[width=0.8\linewidth]{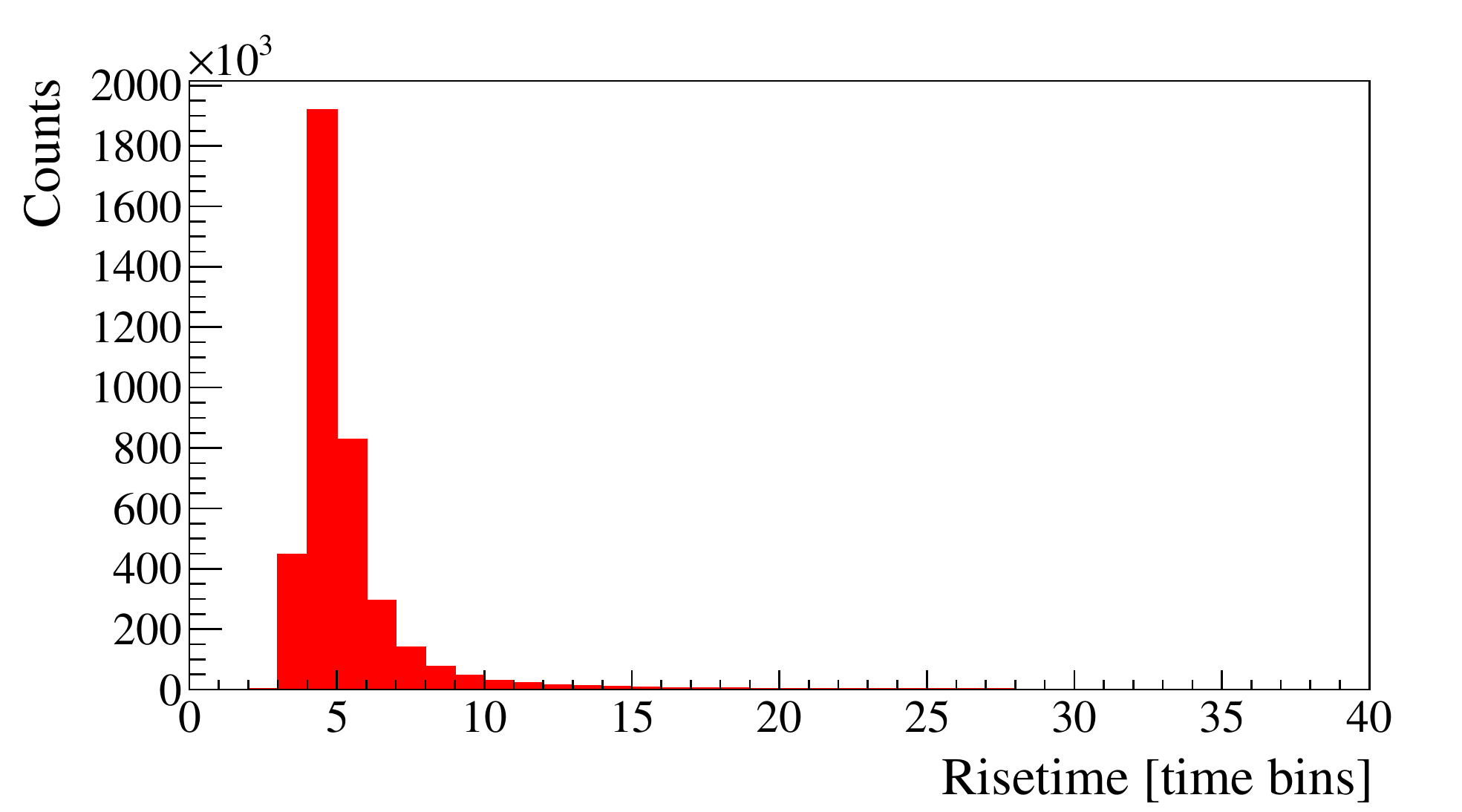}

    \caption{
   Distribution of the signal pulse rise times. The sampling frequency is 5~MHz so each time bin is 0.2~$\mu$s.
    }
    \label{fig:ArRise}
  \end{center}
\end{figure}

The energy of each event is determined from the charge collected on Micromegas readout strips.
More specifically, it is computed by summing the amplitudes of the ADC samples for those channels that passed the programmed threshold.
As mentioned in Section~\ref{sec:MM}, the energy deposited in the detector can be shared between the X and Y channels of Micromegas, which enables the readout plane to produce the 2-D projection of event tracks.
Furthermore, the detector identifies the relative Z (longitudinal direction of the TPC) position by the trigger time of the corresponding signal pulse, as shown on the right of Fig.~\ref{fig:gammaSignal}.

Selection criteria, including number of triggered channels cut, baseline noise cut, signal rise time cut and track position cut, were applied to find the gamma events from $^{241}$Am.
The number of triggered channels is related to the length of a gamma track in the medium.
A drift electron cloud with the size of one Micromegas pixel (zoom-in view in Fig.~\ref{fig:MMStrips}) may trigger both X and Y strips, or only one of them.
In 1~bar Ar-(5\%)isobutane mixture, we chose events with a number of triggered channels between 4 and 29.
For each strip signal, the baseline noise is defined as the RMS (Root Mean Square) of the first 100 sampling points.
The selected events were required to have the baseline RMS less than 3.30~fC.
The rise time cut characterizes the shape of the signal pulse of each strip.
We define the rise time as the time interval between trigger threshold and the pulse peak.
Pulses with short rise times are considered as electronic noise falsely triggered.
And the long rise time often corresponds to the event pile-up, in which case the energy will be reconstructed incorrectly.
The accepted pulse rise time should be within 0.4 and 1.6~$\mu$s.
The track position cut was applied to avoid broken strips of Micromegas.
Strips might have no signal at all due to a connection problem between the signal cable and the electronics, or issues with the Micromegas itself.
The cut selected tracks whose projections lie near the calibration source projection on the readout plane.

\begin{figure}[tb]
  \begin{center}
    \includegraphics[width=0.99\linewidth]{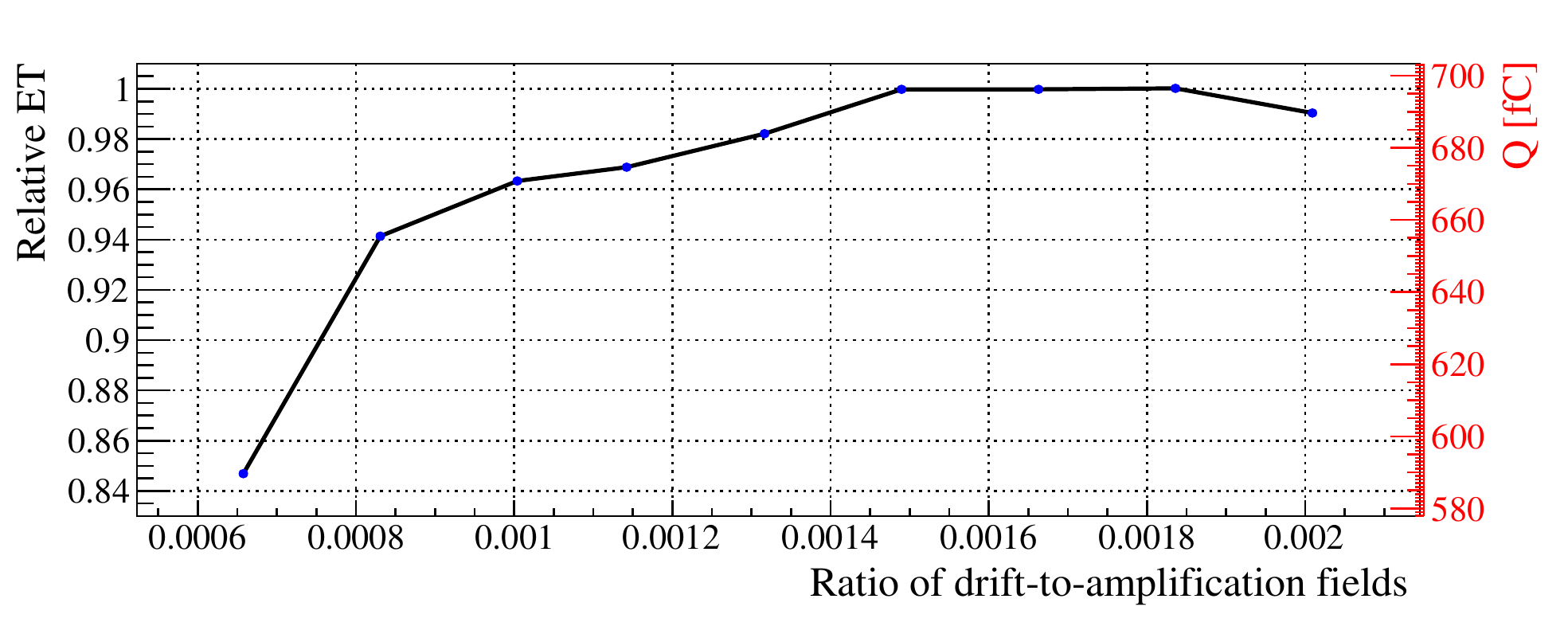}

    \caption{
   Dependence of the electron transmission of Micromegas on drift field in 1~bar Ar-(5\%)isobutane gas mixture.
   The right axis shows the charges collected by the readout strips, characterized by the position of the fitted 13.9~keV peak in the spectrum.
   The left axis shows the relative ET, which is the collected charges normalized to the maximum value measured during the drift field scan.
    }
    \label{fig:ET}
  \end{center}
\end{figure}

The electron transmission (ET) is the ratio of the electrons arriving the amplification region to the primary ionized electrons, indicating the transparency of the TPC.
It is mainly affected by the drift field shape, as well as other factors such as impurity attachment or diffusion effect~\cite{Chefdeville:2009zz}.
In practice the ET was measured for a fixed amplification field of 74~kV/cm and the 10 drift fields listed in Table~\ref{data-statistic}.
Assuming that the gain of Micromegas stays stable when the amplification field is fixed, charges collected by the readout strips are proportional to the electron transmission.
In Fig.~\ref{fig:ET} we show charges collected by the readout strips for 13.9~keV events and the relative ET value, which is  normalized to the highest point in the curve.
Drift-amplification field ratio was chosen as the X-axis variable since it characterized the drift field shape.

The ET curve measured in 1~bar Ar-(5\%)isobutane is shown in Fig.~\ref{fig:ET}.
The electron transmission rises as the drift-amplification ratio increases, plateaus at ratios between 1.49$\times10^{-3}$ and 1.83$\times10^{-3}$ (corresponding to drift fields between 0.11 and 0.136~kV/cm).
For higher drift fields, the field lines coming out of the Micromegas pitch holes contract to a relatively smaller region in the active volume, rendering the decrease in electron collection efficiency of the readout plane.
This effect may explain a slight decrease of the relative ET at the last data point of the curve.
Similar results were reported in~\cite{Iguaz:2012ur,Dafni:2009jb} with different types of gas and detector setups.
The energy spectrum shown later was measured at 0.123~kV/cm (corresponding to a drift-amplification field ratio of $1.66\times10^{-3}$), which is in the ET plateau range.

The gain of the detector $G$ is estimated by:
\begin{equation}
G=\frac{Q}{E/W}
\end{equation}
where $Q$ is the amplified charges collected by the electronics, $E$ is the energy of the corresponding incident particle and $W$ is the average work per pair production.
In Ar-(5\%)isobutane mixture the value of $W$ is taken as 26.2~eV, estimated from the weighted harmonic mean of the $W$ values of argon and isobutane~\cite{Smirnov:2005yi}.
For drift field of 0.123~kV/cm the gain is about 7$\times10^{3}$, calibrated by the 17.8~keV peak in the spectrum.

%[ALTERNATIVE]
%The gain of the detector is characterized by the ratio of the charges collected by the electronics to the corresponding primary ionized electrons $Q_{0}$, where $Q_{0}$ can be estimated by the energy of the incident particle over the total mean work per pair production $W$.
%In Ar-(5\%)isobutane mixture we take the $W$ factor as 26.2~eV~\cite{Smirnov:2005yi}.
%For drift field of 0.123~kV/cm the gain is about 7000, calibrated by the 17.8~keV events.

\begin{figure}[tb]
  \begin{center}
    \includegraphics[width=0.99\linewidth]{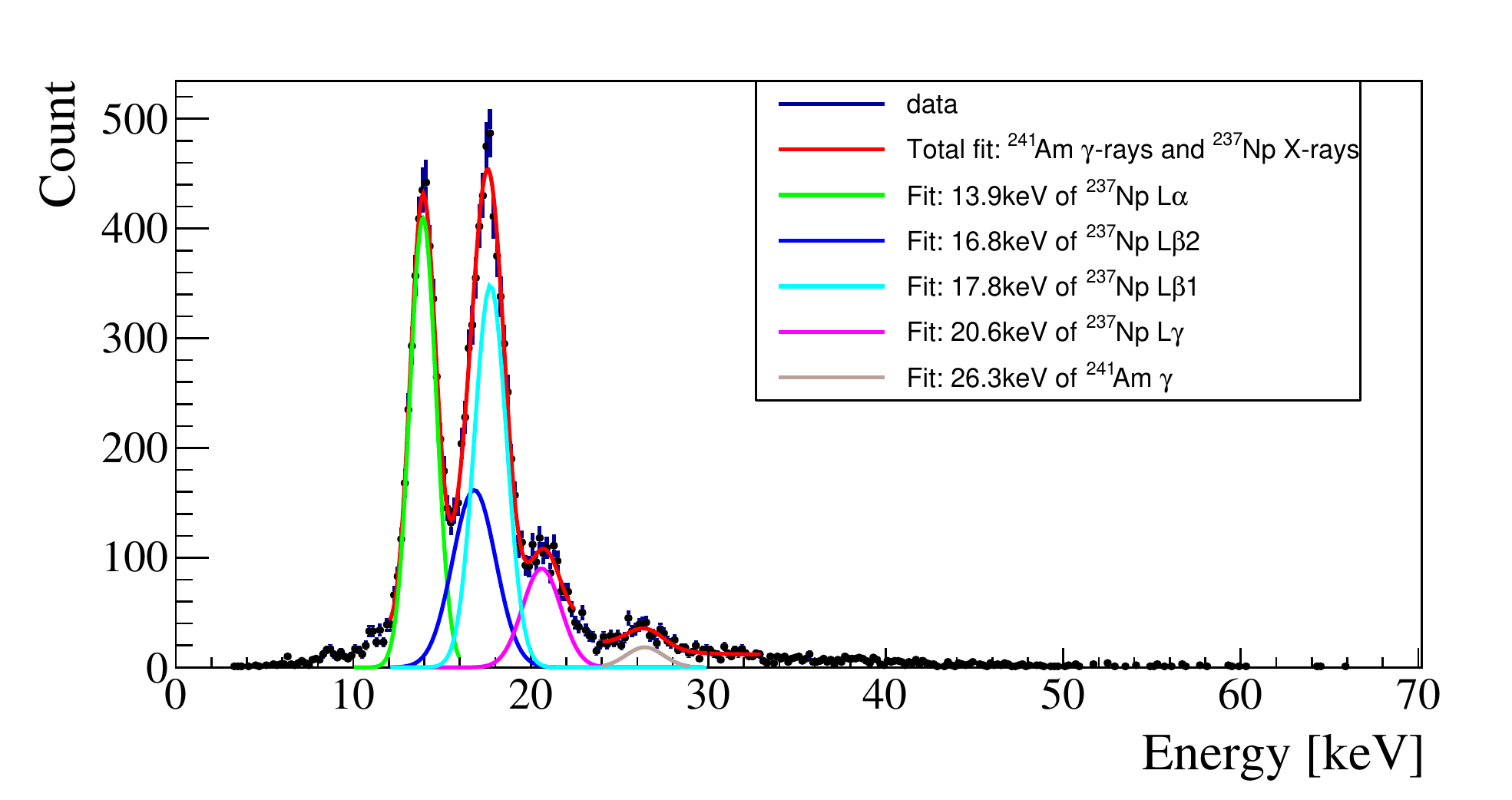}
    \caption{
   Energy spectrum of the $^{241}$Am events in 1 bar Ar-(5\%)isobutane. The fits implemented for each peak are depicted. Their energy resolution (\%FWHM) values: 12.5\% FWHM at 13.9~keV, 11.9\% at 16.8~keV, 11.2\% at 17.8~keV and 9.3\% at 26.3~keV.
    }
    \label{fig:Am241ArSpec}
  \end{center}
\end{figure}

The calibration spectrum after applying the selection criteria is shown in Fig.~\ref{fig:Am241ArSpec}.
The energy spectrum obtained is well consistent with the simulation results (Fig.~\ref{fig:simulation} (Top)).
The energy resolution for the Micromegas in 1 bar Ar-(5\%)isobutane is 12.5\% FWHM at the 13.9~keV peak, 11.9\% at the 17.8~keV, 11.2\% at the 20.6~keV peak and 9.3\% at the 26.3~keV peak.
The energy resolution at 16.8~keV is 17.1\%, smeared by the two smaller peaks near-by (see Fig.~\ref{fig:simulation} (Top Left)).
No visible 59.5~keV gamma peak was detected due to the long attenuation length of the gamma.

\subsection{Data with 5~bar Xe-(1\%)TMA mixture}

\begin{figure}[tb]
  \begin{center}
    \includegraphics[width=0.99\linewidth]{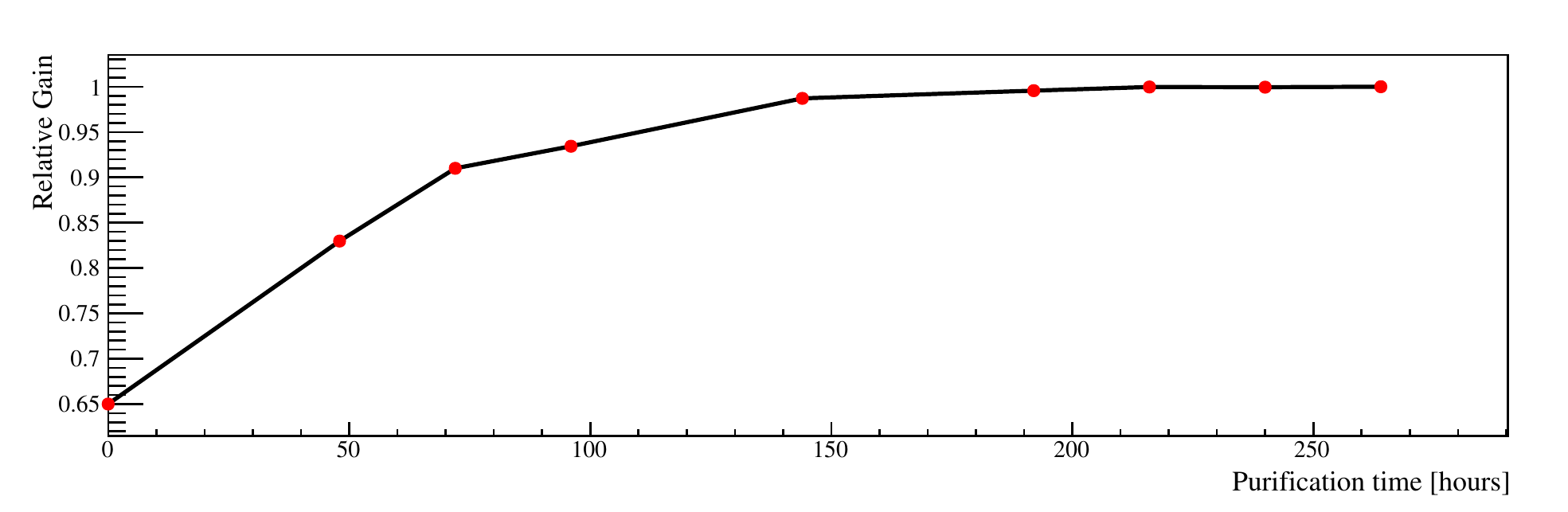}

    \caption{
   The detector gain curve in 5~bar Xe-(1\%)TMA gas mixture, evolving with the purification time. Each data point represents the detector gain  taken at a drift voltage of $-$26~kV and a Micromegas mesh voltage of $-$440~V.
    }
    \label{fig:XePurification}
  \end{center}
\end{figure}

We also took calibration data with 5~bar Xe-(1\%)TMA gas mixture~\cite{Cebrian:2012sp}, which will be the chosen gas mixture for the PandaX-III experiment.
Extra attention was paid to filling the gas into the TPC, as mentioned in an earlier section.
After filling, the gas mixture was circulated continuously at an average flow rate of 3~L/min and purified with a room temperature getter to optimize the detector performance.
The operating HV for the Micromegas module was $-$440~V, corresponding to an amplification field of 88~kV/cm.
The voltage of the cathode was set to the highest possible value to maximize drift electron survival probability.
The results shown in this section were taken at drift voltage of $-$26~kV, which corresponds to a drift field of 333~V/cm.
Detector performance improved with gas purification.
We plot the relative gain curve vs. gas purification time after filling in Fig.~\ref{fig:XePurification}.
The gain values have been normalized by the maximal value we could reach, after which the gain remained stable with continuous gas purification.

\begin{figure}[tb]
  \begin{center}
    \includegraphics[width=\linewidth]{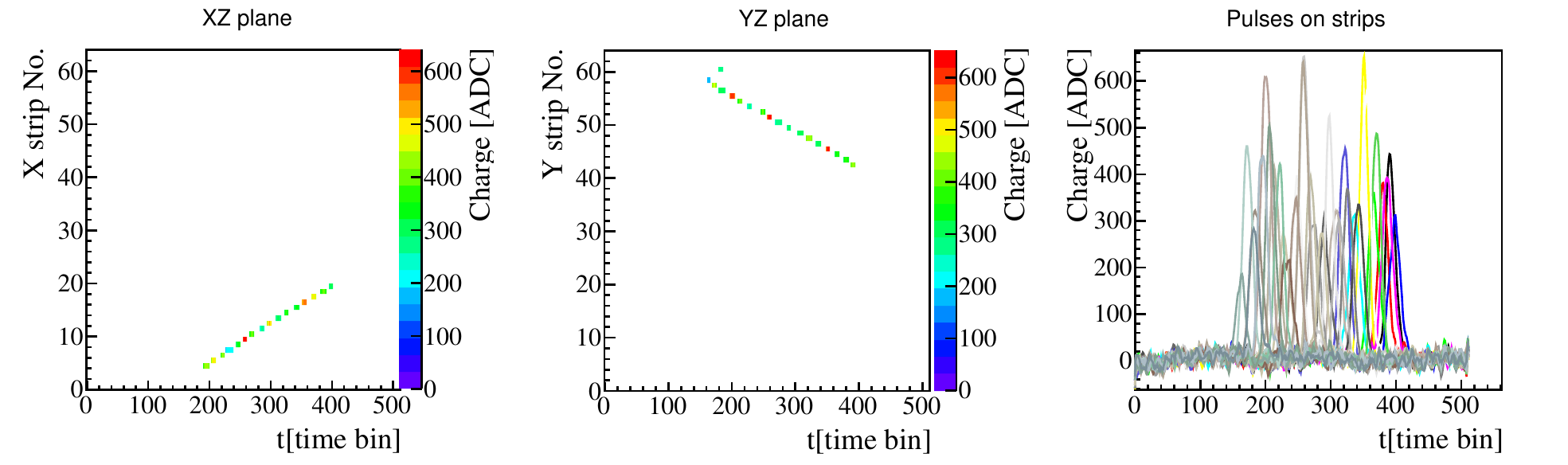}

    \caption{
   A muon event going through the TPC in 5~bar Xe-(1\%)TMA gas mixture. The projected muon tracks on XZ plane (Left) and YZ plane (Middle) are shown.
   Pulses collected on each triggered strip are shown in the right panel, where each color indicates one channel.
    }
    \label{fig:XeMuon}
  \end{center}
\end{figure}

\begin{figure}[tb]
  \begin{center}
    \includegraphics[width=0.6\linewidth]{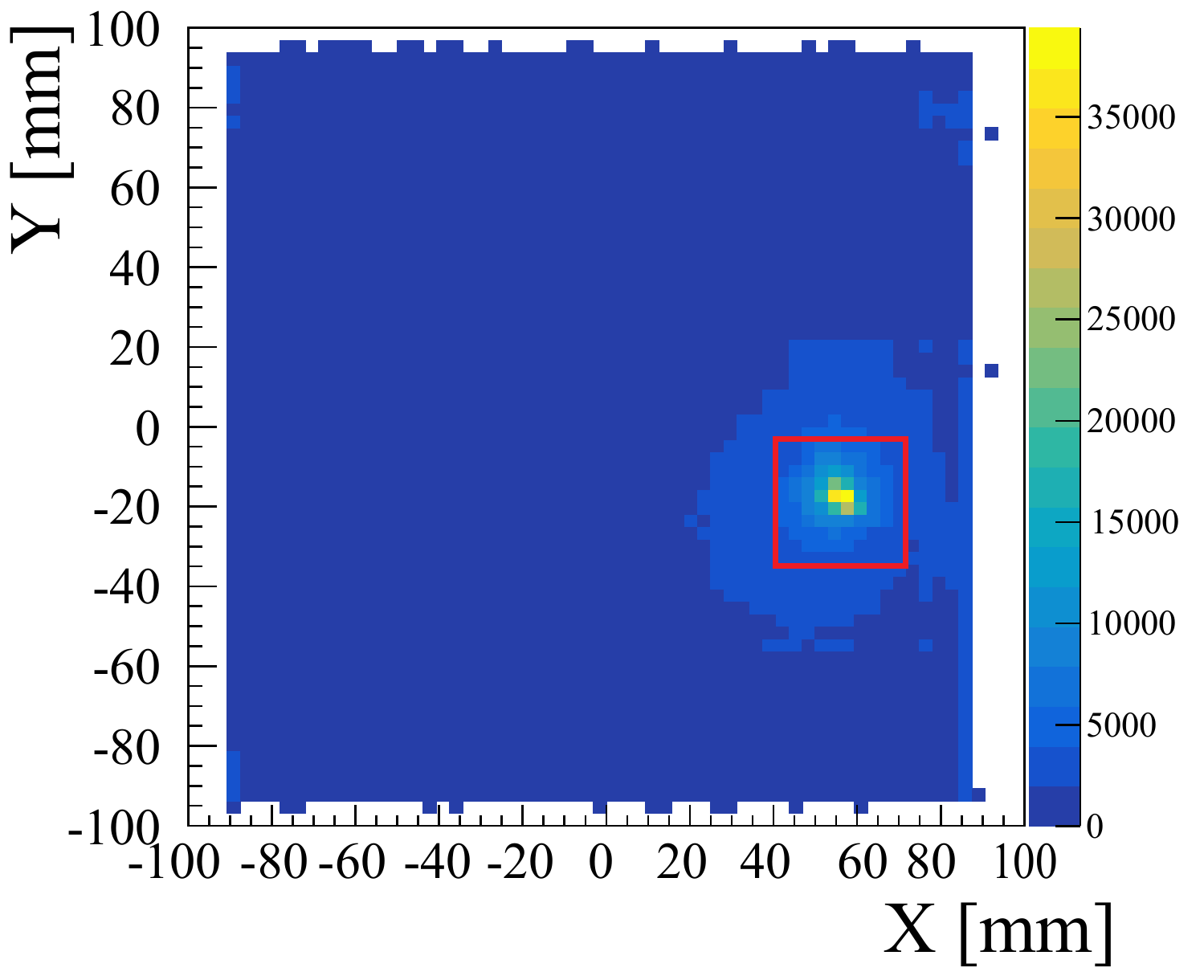}
    \caption{
   $x$-$y$ transverse position distribution of $\gamma$ events from the $^{241}$Am source, obtained on the TPC readout plane (20$\times$20 cm$^2$ Micromegas) under a 3~mm binning, in 5~bar Xe-(1\%)TMA gas mixture. The events with track center points in the red square were chosen by the positioning cut.
    }
    \label{fig:XeHitmap}
  \end{center}
\end{figure}

A typical track of a muon going through the TPC is plotted in Fig.~\ref{fig:XeMuon}.
The track projection on the XZ plane is shown on the left of the figure, and the YZ projection in the middle.
Muon tracks highlight the unique tracking feature of the gaseous TPC.
Furthermore, we are developing analysis techniques to characterize the detector response at different Z positions with the help of penetrating muons.
In Fig.~\ref{fig:XeHitmap}, we show the 2-D histogram of the centers of collected tracks.
The highlighted cluster on the right overlays with the expected position of the $^{241}$Am source and demonstrates the \emph{imaging} capability of the TPC.

\begin{figure}[tb]
  \begin{center}
    \includegraphics[width=0.99\linewidth]{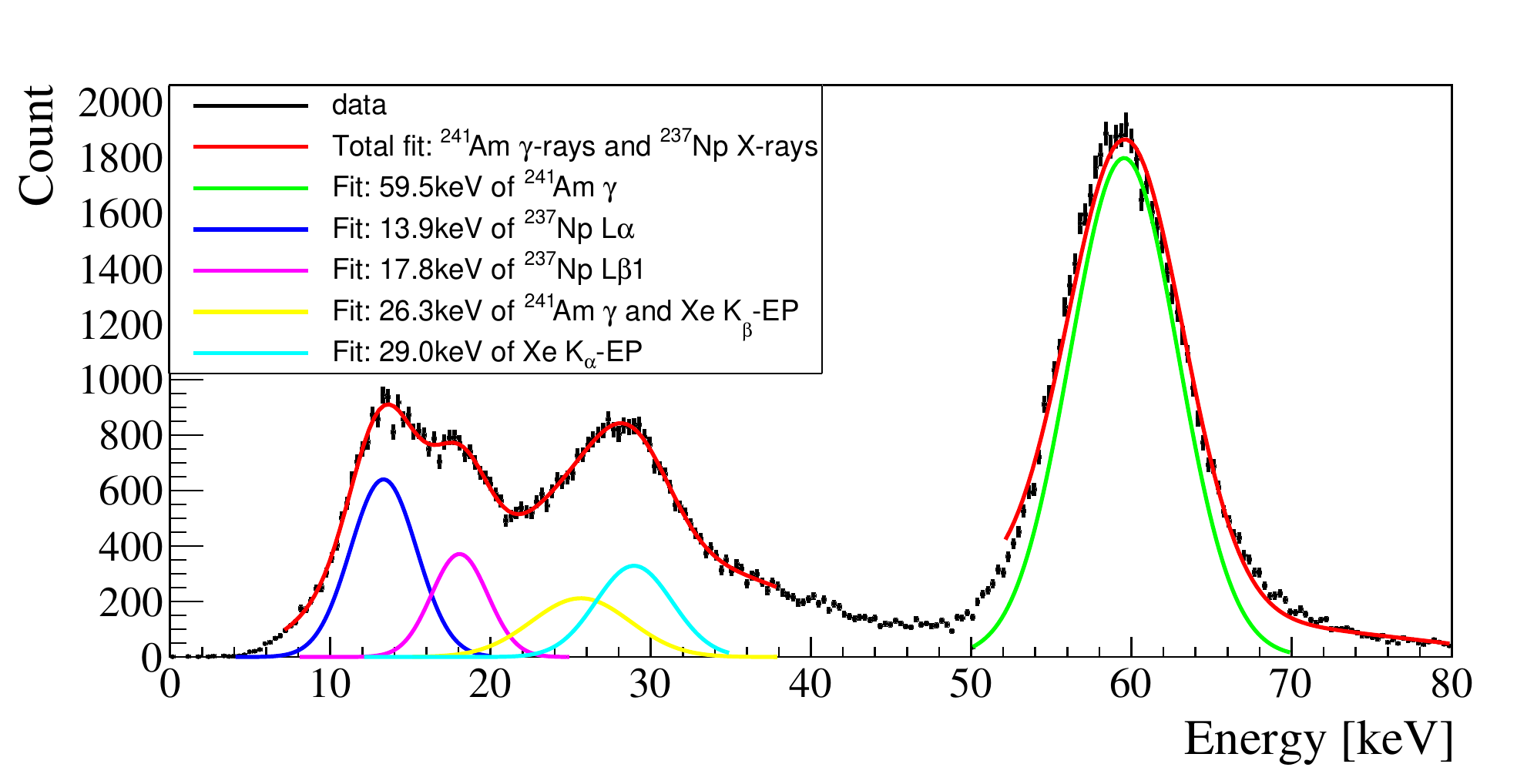}
    \caption{
   Energy spectrum of the $^{241}$Am events in 5 bar Xe-(1\%)TMA gas mixture with all selection criteria applied. The fits implemented for each peak are depicted. Their energy resolution (\%FWHM) values: 14.1\% at 59.5~keV, 19.3\% at 29.0~keV, 28.2\% at 26.3~keV, 22.7\% at 17.8~keV, and 36.5\% at 13.9~keV.
    }
    \label{fig:XeFit}
  \end{center}
\end{figure}

\begin{figure}[tb]
  \begin{center}
    \includegraphics[width=0.6\linewidth]{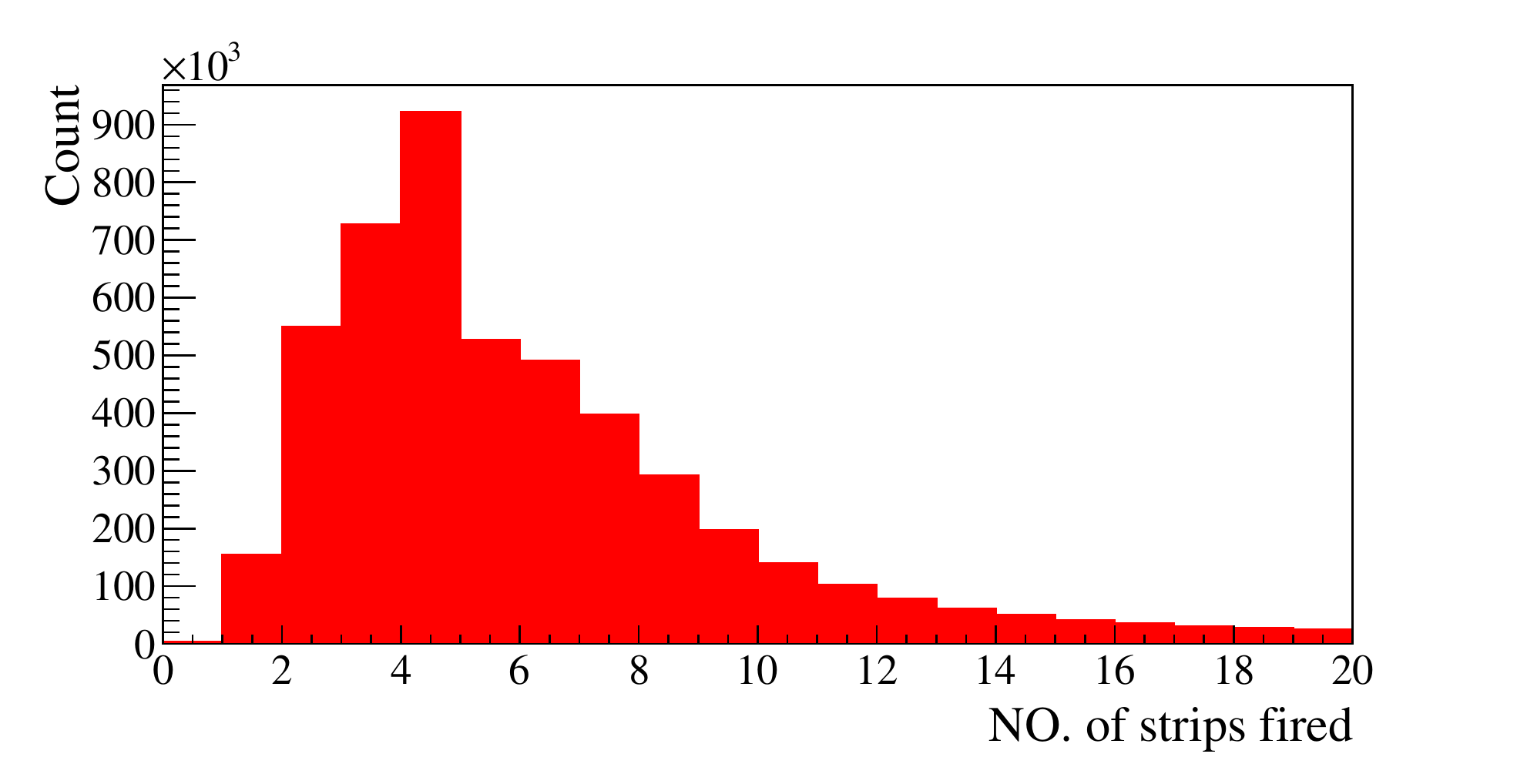}

    \caption{
   The distribution for the fired strips per event in 5~bar Xe-(1\%)TMA gas mixture. Events which fired from 4 to 7 strips were selected in the spectrum shown in Fig.~\ref{fig:XeFit}.
    }
    \label{fig:XeSignal}
  \end{center}
\end{figure}

The calibration spectrum of $^{241}$Am after applying event selection cuts is shown in Fig.~\ref{fig:XeFit}.
The applied cuts are similar to those in the Ar-(5\%)isobutane runs.
For example, Fig.~\ref{fig:XeSignal} shows the distribution of the triggered channel numbers per event and we selected events with the value between 4 and 7.
Multiple strips can be easily triggered due to the interweaving chessboard strip pattern of Micromegas.
The diffusion effect of drift electrons can further increase the triggered channel numbers of some events.
In the track position cut, we chose events with track center points within the 3$\times$3~cm$^2$ red square in Fig.~\ref{fig:XeHitmap}.
27.2\% of the events were selected by the positioning cut.

After all the cuts, The calibration spectrum was fitted with multiple Gaussian peaks.
The spectrum is consistent with our simulation result, where the 59.5~keV gamma peak from $^{241}$Am dominates as expected.
The energy resolution (\%FWHM) is 14.1\% at 59.5~keV, 19.3\% at 29.0~keV, 28.2\% at 26.3~keV, 22.7\% at 17.8~keV, and 36.5\% at 13.9~keV.
In Xe-(1\%)TMA mixture the $W$ value is taken as 22.1~eV~\cite{Smirnov:2005yi} and the detector gain is about 3$\times10^{2}$, calibrated by the 59.5~keV peak.
If we do not include the positioning cut, the energy resolution deteriorates to 15.4\% at 59.5~keV.

\begin{figure}[tb]
  \begin{center}
    \includegraphics[width=0.6\linewidth]{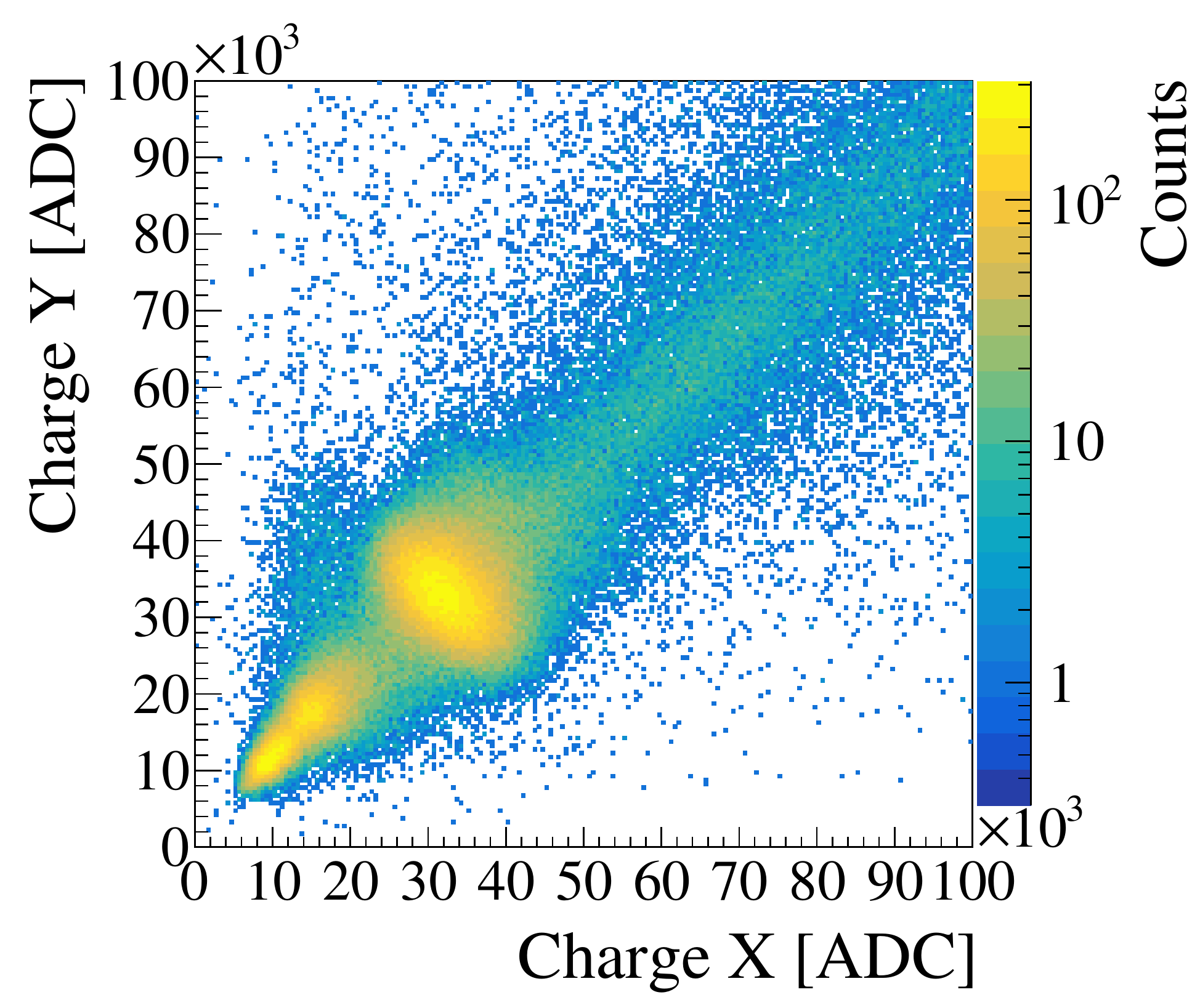}
    \caption{
   The 2D distribution of the charge collected by X strips versus the charge collected by Y strips for events in the Xe-(1\%)TMA mixture.
   Event selection cuts are identical to the spectrum shown in Fig.~\ref{fig:XeFit}.
    }
    \label{fig:Xesharing}
  \end{center}
\end{figure}

Fig.~\ref{fig:Xesharing} demonstrates the charge sharing between X and Y strips.
The charge collected by X strips and the charge collected by Y strips in each event were counted and the histogram was generated on an event by event basis.
Event selection cuts used here are identical to the spectrum shown in Fig.~\ref{fig:XeFit}.
In the figure, gamma peaks such as the 59.5 keV peak are visible.
At lower energy peak region, the charge sharing between X and Y strips are not symmetric, since the source position likely favors one strip on one orientation (in our case, a Y strip) significantly.
At higher energy, the charge sharing does become more symmetrical.

\section{Conclusions and Outlook}
\label{sec:outlook}

In this paper we reported the design of a high pressure TPC that serves as a technology demonstrator for the PandaX-III rare event search experiment. We described the experimental setup and show initial commissioning results with a low energy gamma calibration source.
We used Microbulk Micromegas modules with strip channels for electron amplification and readout.
Our setup is one of the first applications of such Micromegas.
Each of the modules has an active area 20$\times$20 cm$^2$ and 7 modules have been installed in our TPC.
All 128 channels per Micromegas modules are connected to a commercial electronics system based on AGET chips.
Digitized event pulses were recorded at a sampling rate of 5~MHz during data taking.
Offline event construction calculated the energy of the events and the relative triggering time of the fired strips.
Two two-dimensional tracks can then be reconstructed with energy deposition information, which can be used in future NLDBD experiments for background suppression and signal identification.

After validating its leak tightness and HV stability, the TPC was calibrated with a $^{241}$Am source in 1~bar Ar-(5\%)isobutane and 5~bar Xe-(1\%)TMA mixtures, among other gas mediums and different pressure settings.
During the calibration runs, only the central Micromegas module was used.
In the Ar-(5\%)isobutane mixture, we achieved modest energy resolutions of 12.5\% FWHM at 13.9keV and 11.9\% at 17.8keV.
The detector amplification was estimated at 7$\times10^{3}$ for a drift field of 123~V/cm and an amplification field of 74~kV/cm.
In the Xe-(1\%)TMA mixture, the energy resolution at 59.5~keV is 14.1\%.
The detector amplification was estimated at 3$\times10^{2}$ for a drift field of 333~V/cm and an amplification field of 88~kV/cm.
The initial commissioning results are a critical first step for understanding the performance of Microbulk Micromegas with strip readout and its potential for future rare event searches such as NLDBD experiments.

Fine tuning of the TPC performance and data taking with 7 Micromegas modules are on-going.
The commissioning and operation of a detector with such a large Microbulk Micromegas readout area will be an important technical achievement.
Future runs will emphasize on data taking with 10~bar Xe-(1\%)TMA mixture, which is the baseline gas mixture for the PandaX-III project.
With higher pressure, we will also calibrate the TPC by using high energy gamma sources such as $^{137}$Cs and $^{208}$Tl.

\acknowledgments

This work was supported by grant No. 11775142 from the National Natural Science Foundation of China, the National Key Programme for Research and Development (NKPRD) Grant \#2016YFA0400300 from the Ministry of Science and Technology, China.
This project has been supported by a 985-III grant from Shanghai Jiao Tong University, the Key Laboratory for Particle Physics, Astrophysics and Cosmology, Ministry of Education, China, and Chinese
Academy of Sciences Center for Excellence in Particle Physics (CCEPP). We also acknowledge the tremendous contribution made by Xiang Xiao during the design and construction of the prototype TPC.

\bibliographystyle{unsrt}
\bibliography{P3_Prototype_1MM}
\end{document}